\documentclass[%
reprint,
superscriptaddress,
amsmath,amssymb,
aps,
pra,
]{revtex4-1}
\usepackage[utf8]{inputenc}
\usepackage[english]{babel}
\usepackage[T1]{fontenc}
\usepackage{amsfonts}
\usepackage{amsmath}
\usepackage{amsthm}
\usepackage{amssymb}
\usepackage{mathtools}
\usepackage{hyperref}

\usepackage{color,soul}
\usepackage[normalem]{ulem}
\usepackage{braket}

\usepackage[caption=false]{subfig}

\newcommand{\tr}{\mathrm{Tr}}
\newcommand{\calp}{\mathcal{P}}

\newcommand{\ketbra}[2]{\ket{#1} \hspace{-.5ex} \bra{#2}}

\begin{document}
\title{Semi-device-independent characterization of quantum measurements under a minimum overlap assumption}

\author{Weixu Shi}
\affiliation{Department of Electronic Science, National University of Defense Technology, 410073 Changsha, China}
\affiliation{Département de Physique Appliquée, Université de Genève, 1211 Genève, Switzerland}
\author{Yu Cai}
\affiliation{Département de Physique Appliquée, Université de Genève, 1211 Genève, Switzerland}
\author{Jonatan Bohr Brask}
\affiliation{Department of Physics, Technical University of Denmark, Fysikvej, Kongens Lyngby 2800, Denmark}
\author{Hugo Zbinden}
\affiliation{Département de Physique Appliquée, Université de Genève, 1211 Genève, Switzerland}
\author{Nicolas Brunner}
\affiliation{Département de Physique Appliquée, Université de Genève, 1211 Genève, Switzerland}
\date{\today}
\begin{abstract}
Recently, a novel framework for semi-device-independent quantum prepare-and-measure protocols has been proposed, based on the assumption of a limited distinguishability between the prepared quantum states. Here, we discuss the problem of characterizing an unknown quantum measurement device in this setting. We present several methods to attack the problem. Considering the simplest scenario of two preparations with lower bounded overlap, we show that genuine 3-outcome POVMs can be certified, even in the presence of noise. Moreover, we show that the optimal POVM for performing unambiguous state discrimination can be self-tested. 
\end{abstract}

\maketitle

\section{Introduction}

The problem of certifying and characterizing quantum systems is a central problem of quantum information science, in particular towards the development of future quantum technologies. It is desirable to develop certification methods that are highly robust to noise and technical imperfections. 

The device-independent (DI) approach~\cite{barrett_no_2005, acin_device-independent_2007, colbeck_quantum_2009, pironio_random_2010} is of strong interest in this context; see e.g. Ref.~\cite{scarani_black_2014, brunner_bell_2014} for recent reviews. The main feature here is that a quantum system (or device) can be certified with minimal assumptions, without the requirement of using previously calibrated devices. In the fully DI approach, the observation of certain measurement statistics can certify a general property of a quantum system (for instance that a source produces a quantum state that is entangled), and even completely characterize the system, i.e. identify precisely which entangled state is produced). The latter is referred to as ``self-testing'', see e.g. Ref.~\cite{mayers_self_2003, summers_maximal_1987, popescu_generic_1992, mckague_robust_2012, kaniewski_analytic_2016}.

While the fully DI approach is conceptually very elegant and provides the strongest possible form of certification for a quantum system, it is challenging to implement in practice. The main difficulty is that fully DI certification methods require a loophole-free Bell inequality violation. This motivated the development of partially DI methods that can be implemented in simple prepare-and-measure type experiments, which do not involve entanglement. The price to pay for this simplification is that an additional assumption on the system is required. First works in this direction used an assumption on the Hilbert space dimension of the quantum states being prepared \cite{gallego_device-independent_2010, wehner_lower_2008, bowles_certifying_2014,sekatski_certifying_2018}. Self-testing methods have been developed for this setting \cite{tavakoli_self-testing_2018}, for characterizing quantum states and measurements \cite{farkas_self-testing_2019,tavakoli_enabling_2019, tavakoli_self-testing_2018-1,miklin_self-testing_2019,mironowicz_experimentally_2018,Mohan_2019}, as well as for implementing quantum information protocols~\cite{pawlowski_semi-device-independent_2011, li_semi-device-independent_2012, lunghi_self-testing_2015, bourennane_experimental_2004, woodhead_secrecy_2015}. In practice however, the assumption of bounded dimension is not straightforward to justify, as dimension is not a directly measurable quantity. One typically needs to assume that the experimental setup is free of extra side-channels. As this is delicate in practice, one would ideally find other solutions allowing one to discard this assumption. 

This motivates the study of different approaches to the semi-DI setting, using different types of assumptions. Three promising approaches have been recently put forward. First, Ref. \cite{chaves_device-independent_2015} suggested to upper bound the entropy of the quantum message (i.e. the set of prepared quantum states). Then, Ref. \cite{himbeeck_semi-device-independent_2017} proposed an upper bound on the energy of quantum states. Finally, Ref. \cite{brask_megahertz-rate_2017} assumed a lower bound on the overlap between the prepared quantum states. Moreover, Ref.~\cite{wang_characterising_2019} has developed a toolbox to characterize the quantum correlation in the prepare-and-measure scenario under the assumption of overlaps of the quantum states. Clearly, the common feature of all these approaches is placing a bound on how distinguishable the quantum states are from each other. In practice these approaches open new perspectives. Indeed, the energy of an optical source can in principle be directly measured, which provides a good justification for an upper bound on the energy, or a lower bound on the overlap (using, say, the vacuum and weak coherent states). This approach recently led to promising randomness generation protocols \cite{brask_megahertz-rate_2017, rusca_practical_2019}, combining semi-DI security, high rates, and ease of implementation.

Here we explore further the potential of this new approach to the semi-DI setting. In particular, we consider the problem of characterizing an unknown quantum measurement device in a simple prepare-and-measure scenario, which features only two possible preparations and a fixed ternary measurement. We use the assumption of a lower bound on the indistinguishability between the two prepared quantum states, which we formalize for mixed states in terms of lower bounds on the fidelity. This allows us to certify certain properties of the positive-operator valued measure (POVM) that is implemented inside the measurement device. In particular, we show that the observation of certain correlations certifies that the measurement is a genuine 3-outcome POVM. In order to do so, we develop methods to characterize the set of correlations achievable with binary POVMs and classical post-processing. Moreover, we show that a particular genuine 3-outcome POVM, which allows for unambiguous state discrimination \cite{ivanovic_how_1987,dieks_overlap_1988,peres_how_1988}, can be self-tested. Finally, we discuss the robustness to noise of these methods.

\section{Defining the problem} \label{sec:introductionII}

\begin{figure}[t!]
	\centering
	\includegraphics[width=3.5in,trim={7cm 12.55cm 6cm 12.55cm},clip]{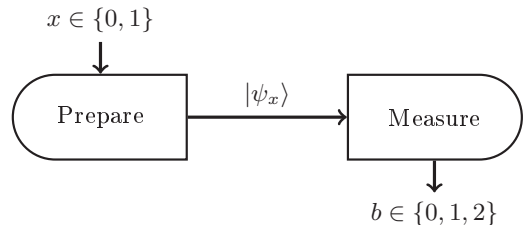}
    \caption{Schematic representation of the scenario considered.}
    \label{fig:scenario}
\end{figure}

We consider the prepare-and-measure setup sketched in Fig.~\ref{fig:scenario}. The preparation device takes a binary input, $x\in\{0,1\}$, and the measurement box performs a fixed measurement (hence no input) resulting in a ternary output, $b\in\{0,1,2\}$. Upon receiving $x$, the preparation device sends a quantum system in an unknown state $\rho_x$ to the measurement device, which performs an unknown POVM on the system. The POVM elements associated to each outcome are noted $M_b$. This results in the following statistics
\begin{align}
p(b|x)=\tr[\rho_x M_b]
\end{align}
the set of which, $\boldsymbol{p}:=\{p(b|x)\}$, is called a \textit{behavior}.

Our goal is to characterize the unknown POVM that is implemented inside the measurement device. This characterization is semi-DI, in the sense that it is based only on the observed behavior, under two assumptions. First, the choice of the input $x$ is independent from the boxes. All the information that the measurement device receives about $x$ comes from the received quantum state $\rho_x$. Hence, in order to make non-trivial statements, we need to limit the amount of information about $x$ that can be retrieved from the states $\rho_x $. This leads to our second assumption, namely that a lower bound on the indistinguishability of the two quantum states. Here we use the fidelity \cite{uhlmann_transition_1976,jozsa_fidelity_1994,Nielsen_QCQ_2011} as a measure of indistinguishability between $\rho_0$ and $\rho_1$. Our assumption reads
\begin{align}\label{assumption}
F(\rho_0,\rho_1)=\tr\sqrt{\rho_0^{1/2}\rho_1\rho_0^{1/2}} \geq \delta.
\end{align}
For the case of two pure states, we have simply that $F = |\braket{\psi_0|\psi_1}|\geq \delta$. Note also that when the two states are identical, then $F(\rho_0,\rho_1)=1$. In the following, without loss of generality, we will restrict our analysis to the case of two pure states. This is because the set of behaviors that is achievable under the above assumption \eqref{assumption} can always be reproduced by using two pure states with the same overlap. To see this, suppose a behavior is produced by two mixed states with $F(\rho_0,\rho_1)=\delta$. According to Uhlmann's theorem, there exists a pair of purifications of $\rho_0$ and $\rho_1$, denoted by $\ket{\phi_0}$ and $\ket{\phi_1}$ respectively, such that their overlap satisfies $|\braket{\phi_0|\phi_1}|=\delta$. Then $p(b|x)=\tr(\rho_x M_b)=\tr(\tr_R(\ketbra{\phi_x}{\phi_x})M_b)=\tr(\ketbra{\phi_x}{\phi_x}M_b\otimes\mathbb{I}_R)$, where $\rho_x=\tr_R(\ketbra{\phi_x}{\phi_x})$ and $R$ is the ancillary system.

Let us discuss the parametrisation of the two pure quantum states. Without loss of generality, we can represent these states in an effective qubit space spanned by the states $\ket{0}$ and $\ket{1}$. Note that we make no assumption on the Hilbert space dimension, but simply use the fact that we can set the reference frame at our convenience. Specifically, we write
\begin{align}\label{eq:states}
	&\ket{\psi_0}=\cos\theta\ket{0}+\sin\theta\ket{1},\nonumber\\
	&\ket{\psi_1}=\cos\theta\ket{0}-\sin\theta\ket{1},
\end{align} with $\delta= \cos2\theta$ and $0 \leq \theta \leq \pi/4$, so that the overlap $\braket{\psi_0|\psi_1}$ is positive and real. Note that since all the behaviors achievable via pairs of quantum states with a larger overlap are included in the behaviors with a smaller overlap (see Appendix 2 of Ref.~\cite{brask_megahertz-rate_2017}), we can take the overlap of the two states to be $\delta$ when characterizing the boundary of the sets of behaviors for an overlap larger than or equal to $\delta$.

With the overlap assumption, the first property of the measurement box to be certified is that it performs a \textit{genuine} 3-outcome POVM, i.e.
a measurement that cannot be decomposed into a convex combination of 2-outcome POVMs. Mathematically, if for all $b$, we can write
\begin{align} \label{eq:strategies1}
	M_b = \sum_j p_j M^j_b, 
\end{align}
where each $\{ M^j_b\}_{b=0,1,2}$ is a valid POVM with $M^j_j = 0$, and $\{p_j\}_{j=0,1,2}$ is a valid probability distribution, then we say that $\left\{ M_b \right\}$ is \textit{not} a genuine 3-outcome POVM. Physically speaking, this means such an $\left\{ M_b \right\}$ could be effectively carried out by applying only 2-outcome POVMs and classical post-processing. 

Let $\calp_3(\delta)$ denote the set of behaviors achievable by 3-outcome POVMs for a fixed overlap $\delta$ (or larger), and $\calp_2(\delta)$ denote the one achievable by a convex combination of 2-outcome POVMs. We should have $\calp_2(\delta) \subsetneq \calp_3(\delta)$ for any $\delta>0$, 
since we know that 2-outcome POVMs are special cases of 3-outcome POVMs. Moreover it has been shown that, in the regime $\delta$ close to one,  behaviors in $\calp_3(\delta)$ can certify more randomness than $\calp_2(\delta) $~\cite{ioannou_upper_2019}.
For completeness, we also introduce another set of behaviors, called the trivial set $\calp_t$. Here the input state is ignored (so $\delta$ is omitted) and the output is generated at random according to some distribution. Such implementations will be called trivial POVMs in the following. Note that this set is different from the set of classical behaviors in Ref.~\cite{himbeeck_semi-device-independent_2017}. Mathematically, $\boldsymbol{p} \in \calp_t$ implies $p(b|0) = p(b|1)$ for all $b$. 
One should note that $\calp_3(\delta)$, $\calp_2(\delta)$ and $\calp_t$ are convex, on which our arguments are based.

Finally, note that the problem of certifying genuine 4-outcome POVMs is not discussed here. While there exist extremal qubit POVM featuring four outcomes, these can never be distinguished from 3-outcome POVMs in the present scenario. This is because we can restrict our analysis to POVMs for which all the elements are in a plane of the Bloch sphere (spanned by the two states \eqref{eq:states}). In this case, extremal POVMs feature only 3 outcomes \cite{ioannou_upper_2019}, hence any behavior can be reproduced via 3-outcome POVMs and classical post-processing. The certification of genuine 4-outcome POVM would require a scenario with 3 preparations with limited distinguishability, a problem which we leave for future research.

%The reason why we did not focus on the behaviors of 4-outcome POVMs here is that they can all be reproduced by mixing 3-outcome POVMs. This can be well explained in a Bloch sphere of $\mathcal{H}_{\psi_0,\psi_1}$. Every POVM acting on $\mathcal{H}_{\psi_0,\psi_1}$ can have a ``projected'' POVM which has its Bloch vectors in the same plane as $\psi_0$ and $\psi_1$ and yields the same behavior. The ``projected'' 4-outcome POVMs can always be decomposed into the convex combination of 3-outcome POVMs (see Lemma 1 of Ref.~\cite{ioannou_how_2018}).

%\dashuline{It is interesting to note an inequivalence between being a genuine 3-outcome POVM and being able to certify randomness generation. There exist genuine 3-outcome POVMs that generate little randomness.} 
%
\section{Results}

In this section, we present different methods for characterizing the sets of behaviors $\calp_3(\delta)$, $\calp_2(\delta)$ and  $\calp_t$. First, we show that the problem of determining whether a certain behavior belongs to $\calp_2(\delta)$ can be cast as a semidefinite program (SDP). Then we determine the boundary of the various sets for a specific class of behaviors. Finally, we show that various properties of the POVM for performing unambiguous state discrimination (USD) can be certified, in particular that the POVM can be self-tested.

\subsection{Semi-definite programs}\label{sec:witness}
Here we show that deciding if a behavior belongs to $\calp_2(\delta)$, or whether it must feature a genuine 3-outcome POVM, can be cast as an SDP. Let $\boldsymbol{p}$ be the behavior of interest, and $\boldsymbol{p}_0$ be arbitrary behavior in $\mathcal{P}_2$. For example, $\boldsymbol{p}_0=\boldsymbol{p}_\mathbb{I}=\{p(b|x)=1/3\}$. Clearly $\boldsymbol{p}_\mathbb{I}\in\calp_t$, thus $\boldsymbol{p}_\mathbb{I}\in\calp_2(\delta)$. Consider the linear combination of these two behaviors $\boldsymbol{p}'=\omega \boldsymbol{p}+(1-\omega)\boldsymbol{p}_0$ with $\omega>0$. Let $\omega^*$ denote the maximal $\omega$ for which $\boldsymbol{p}'\in \calp_2(\delta)$. The quantity $\omega^*$ tells us how far a behavior can go along the direction from $\boldsymbol{p}_0$ to $\boldsymbol{p}$ while staying in $\calp_2(\delta)$. If $\omega^*\geq 1$, it means $\boldsymbol{p} \in \calp_2(\delta)$, otherwise $\boldsymbol{p}\not\in\calp_2(\delta)$.

From Eq. \eqref{eq:strategies1}, we see that the probability to use the $j$-th strategy can be absorbed into the POVM elements, i.e. $\tilde{M}^j_b=p_jM^j_b$.
Then computing $\omega^*$ can be written as the following optimization problem with linear constraints
\begin{align}
\label{eq:primalSDP}
\underset{\tilde{M}^j_b}{\text{maximize}}\quad & \omega \nonumber \\
\text{subject to} \quad 
&\tilde{M}^j_b\succeq0, \;\forall j,b, \nonumber\\
&\sum_{b}\tilde{M}^j_b=\frac{1}{2}\tr[\sum_{b}\tilde{M}^j_b]\mathbb{I},\;\forall j, \nonumber\\
&\sum_{j}\frac{1}{2}\tr[\sum_{b}\tilde{M}^j_b]=1, \\
& \tilde{M}^j_j=0, \;\forall j, \nonumber \\
&\omega p(b|x) + (1-\omega) p_0(b|x) \nonumber \\
& \quad \quad \quad =\tr[\ketbra{\psi_x}{\psi_x}\sum_{j}\tilde{M}^j_b], \quad\forall x,b, \nonumber
\end{align}
The first two constraints stem from the positivity and normalization of $M^j_b$, and the next two constraints guarantee the convex combination of 2-outcome POVMs. The last constraint enforces the reproduction of the behavior.

One way to write the dual problem of the SDP above is 
\begin{align}
\label{eq:dualSDP}
	\underset{H^j,J^j,v_{b|x}}{\text{maximize}} \quad & \eta = \boldsymbol{v}\cdot (\boldsymbol{p}-\boldsymbol{p}_0) \nonumber \\
\text{subject to} \quad 
	&H^j=(H^j)^\dagger, \; J^j=(J^j)^\dagger, \nonumber\\
	&\frac{1}{2}\mathbb{I}+H^j-\frac{1}{2}\tr[H^j]\mathbb{I}+\frac{1}{2}\sum_{xb'}v_{b'|x}p_0(b'|x)\mathbb{I}\nonumber\\
	&\quad\quad\quad -\sum_{x}v_{b|x}\ketbra{\psi_x}{\psi_x}+\delta_{b,j}J^j\succeq0 \quad \forall j,b
\end{align}
where $\boldsymbol{v}\in\mathbb{R}^6$, and $\cdot$ denotes the scalar product.
The details of deriving the dual problem from the primal are given in Appendix~\ref{sec:append_dual}. Any feasible solution to the dual problem gives an upper bound on $\omega^*$ ($\omega^*\leq \frac1\eta$). Let $\eta^*$ denote the maximal $\eta$. Any feasible point $\{H^j,J^j,\boldsymbol{v}\}$ which gives $\eta^*>1$ provides a witness for genuine 3-outcome POVMs, since this feasibility does not depend on $\boldsymbol{p}$. For such a feasible point, for any behavior $\boldsymbol{q}$ that violates the inequality
\begin{align*}
	\boldsymbol{v}\cdot (\boldsymbol{q}-\boldsymbol{p}_0) \leq 1,
\end{align*}
 we have $\boldsymbol{q}\not\in \calp_2(\delta)$.
These SDP methods will be used in the next section on specific examples.

\subsection{Analytical characterization of boundary} 
\label{sec:analytical}
Another approach to distinguishing $\calp_2(\delta)$ and $\calp_3(\delta)$ is to characterize their respective boundaries. Even though determining the boundary of quantum correlation in general is challenging, we are able to characterize them for a specific class of behaviors. 

For convenience, we write the vector $\boldsymbol{p}$ of a given behavior in the form
\begin{align}
\label{eq:distribution_notation}
\begin{pmatrix}
p(0|0)      & p(1|0)      & p(2|0)    \\
p(0|1)      & p(1|1)      & p(2|1)      \\
\end{pmatrix}.
\end{align}

We defined $\mathcal{P}_\text{sym}(\delta)$ to be the subset of behaviors in $\mathcal{P}_3(\delta)$ that are invariant to the input-output relabeling 
$$\Pi: \begin{pmatrix}
a      & b      & c    \\
d      & e      & f      \\
\end{pmatrix} \mapsto 
\begin{pmatrix}
e      & d      & f    \\
b      & a      & c      \\
\end{pmatrix}.$$
Notice that the behaviors in $\mathcal{P}_\text{sym}(\delta)$ have the form \begin{align}
&\boldsymbol{p}(X,Y) =\nonumber\\  
&\quad X
\begin{pmatrix}
1      & 0      & 0    \\
0      & 1      & 0      \\
\end{pmatrix}
+Y
\begin{pmatrix}
0      & 1      & 0    \\
1      & 0      & 0      \\
\end{pmatrix}
+(1-X-Y)
\begin{pmatrix}
0      & 0      & 1    \\
0      & 0      & 1      \\
\end{pmatrix}.
\label{eq:slice}
\end{align}
From this we can see $\mathcal{P}_\text{sym}(\delta)$ is in the slice $S$ in $\mathbb{R}^6$. Hence the behaviors in $\mathcal{P}_\text{sym}(\delta)$ can be parameterized by $$X=\frac{1}{2}\left(p(0|0)+p(1|1)\right)$$ and 
$$Y =\frac{1}{2}\left(p(0|1)+p(1|0)\right).$$

Now we introduce a map, $T$, from a general behavior to a behavior in S: $T(\boldsymbol{p})=1/2(\boldsymbol{p}+\Pi(\boldsymbol{p}))$. Apparently, $T(\mathcal{P}_\text{sym}(\delta))=\mathcal{P}_\text{sym}(\delta)$. Our interest lies in the difference of $T(\calp_2(\delta))$ and $T(\calp_3(\delta))$ in the slice $S$.

Notice that $\Pi$ does not change the number of genuine measurement outcomes to reproduce a behavior because it is just relabeling the inputs and outputs. Hence for any $\boldsymbol{p}\in\mathcal{P}_k(\delta)$, $\Pi(\boldsymbol{p})\in\mathcal{P}_k(\delta)$. From the linearity of $T$ and the convexity of $\mathcal{P}_k(\delta)$, we conclude that $T(\boldsymbol{p})\in\mathcal{P}_k(\delta)$, namely, $\mathcal{P}_k(\delta)$ is closed under $T$.

To characterize $T(\calp_2(\delta))$, we can focus on the extremal points of it because of linearity of $T$ and convexity of $\mathcal{P}_2(\delta)$. The extremal points are yielded by projective 2-outcome POVMs and the trivial POVMs. First note that $\calp_t=\calp_3(\delta=1)=\calp_2(\delta=1)$, and it constitutes the line segment connecting $(0,0)$ and $(1/2,1/2)$. Moreover, $\calp_3(\delta=0)=\calp_2(\delta=0)$ corresponds to the full triangle with extremal points $(0,0)$, $(0,1)$ and $(1,0)$. This is because for perfectly distinguishable states, any statistics can be produced by the measurements. As in Eq.~\eqref{eq:strategies1}, we have three 2-outcome strategies, written as $\{0, K_1, \mathbb{I}-K_1\}$, $\{K_2,0,\mathbb{I}-K_2\}$, and $\{K_3,\mathbb{I}-K_3,0\}$, where $K_i$ denotes one of the elements of the $i$th 2-outcome measurement. For convenience, we consider projective 2-outcome POVMs and trivial POVMs separately.

Strategies $\{0, K_1, \mathbb{I}-K_1\}$ and $\{K_2,0,\mathbb{I}-K_2\}$ yield the same ellipse
\begin{equation}\label{eq:P2ellipse}
	\frac{4(X+Y-1/2)^2}{\delta^2}+\frac{4(X-Y)^2}{1-\delta^2}=1.
\end{equation}
Strategy $\{K_3,\mathbb{I}-K_3,0\}$ contributes to the line segment of $X+Y=1$ between the points
\begin{align}\label{eq:vertice1}
	(\frac{1-\sqrt{1-\delta^2}}{2},\frac{1+\sqrt{1-\delta^2}}{2})
\end{align} and
\begin{align}\label{eq:vertice2}
	(\frac{1+\sqrt{1-\delta^2}}{2},\frac{1-\sqrt{1-\delta^2}}{2}).
\end{align}
The details to derive these are given in Appendix \ref{sec:append_Characterzation of slice}.

\begin{figure}[htbp]
	\centering
	\subfloat[]{\label{fig:slice_geo}
		\includegraphics[width=2.7in,trim={4cm 8.5cm 4cm 9cm},clip]{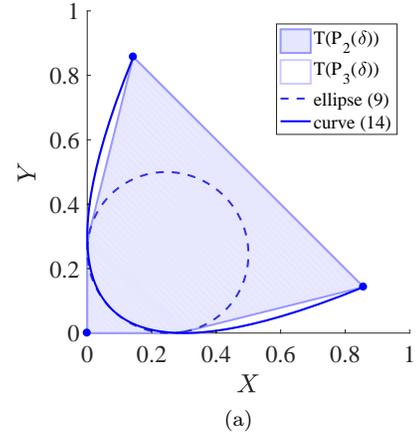}
	}

	\subfloat[]{
		\includegraphics[width=2.5in,trim={4cm 9.5cm 4cm 9.5cm},clip]{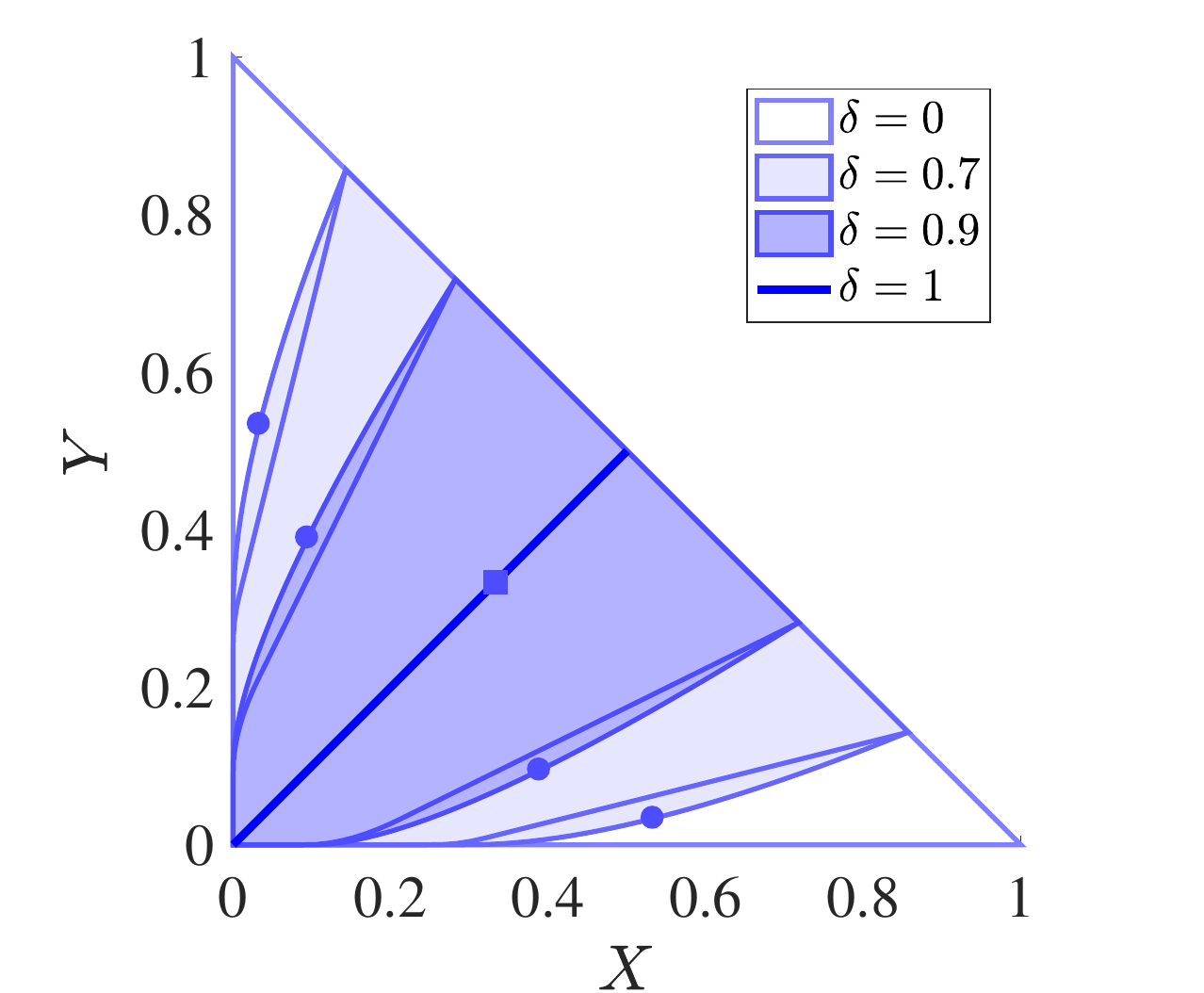}
		\label{fig:slice_delta}
	}
	\caption{(a) Geometrical representation of the sets of $T(\calp_2(\delta))$ and $T(\calp_3(\delta))$ in $S$ at $\delta=0.7$.
	(b) $T(\calp_2(\delta))$ and $T(\calp_3(\delta))$ of $\delta=0$, $0.7$, $0.9$ and $1$. The regions of smaller overlaps cover the regions of larger overlaps. The behaviors corresponding to the genuine 3-outcome that is most robust to noise ($M_\text{rob}$) in Sec.~\ref{subsubsec:robust3-outcome} are marked out here using circles. The square represents the uniformly distributed behavior.}
\end{figure}

Hence, $T(\mathcal{P}_2(\delta))$ is the convex hull of the points $(0,0)$, Eq.~\eqref{eq:vertice1}, Eq.~\eqref{eq:vertice2}, and the ellipse~\eqref{eq:P2ellipse}. $T(\boldsymbol{p})\notin T(\calp_2(\delta))$ certifies a genuine 3-outcome POVM. Note that this is a nonlinear witness, contrary to the witnesses derived from SDP which are linear; see Sec.~\ref{sec:witness}.

% characterization of p3 in this projection

%To characterize $T(\mathcal{P}_3(\delta))$, we show that it is sufficient to consider only a subset $\mathcal{P}_\text{sym}(\delta)$ of $\mathcal{P}_3(\delta)$ which features symmetry in the probabilities. Let $\boldsymbol{p}$ be written
%\begin{align}
%\label{eq:distribution_notation}
%\begin{pmatrix}
%p(0|0)      & p(1|0)      & p(2|0)    \\
%p(0|1)      & p(1|1)      & p(2|1)      \\
%\end{pmatrix}.
%\end{align}

\begin{figure}[t!] 
	\centering
	\includegraphics[width=1.4in,trim={9cm 12.2cm 9cm 12.2cm},clip]{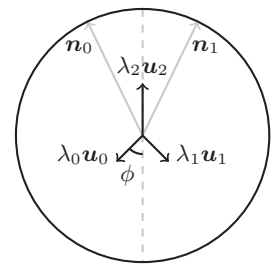}
%	\centerline{
%		\begin{tikzpicture}[thick, global scale = 1.7]
%		\draw[->,color=gray!50] (0,0) -- (0.4359,0.9); 
%		\draw[->,color=gray!50] (0,0) -- (-0.4359,0.9); 
%		\draw[dashed,color=gray!50] (0,1) -- (0,-1);
%		\draw[->] (0,0) -- (0.2071,-0.2071); 
%		\draw[->] (0,0) -- (-0.2071,-0.2071); 
%		\draw[->] (0,0) -- (0,0.4142); 
%		\draw (0,0) circle (1);  	
%		\draw (-0.1036,-0.1036) arc (225:270:0.1464); % arc
%		\tikzstyle{every node}=[font=\small];
%		\node (phi) at (-0.12,-0.31){$\phi$};% text
%		\node (u0) at (-0.47,-0.15){$\lambda_0\boldsymbol{u}_0$};
%		\node (u1) at (0.47,-0.15){$\lambda_1\boldsymbol{u}_1$};
%		\node (u2) at (0,0.55){$\lambda_2\boldsymbol{u}_2$};
%		\node (n0)	at (-0.5,0.7){$\boldsymbol{n}_0$};
%		\node (n1)	at (0.5,0.7){$\boldsymbol{n}_1$};				
%		\end{tikzpicture}	
%	}
	\caption{Schematic representation of the 3-outcome POVMs with symmetrical Bloch vectors. The first POVM element has a Bloch vector pointing in the $z$ direction, i.e. intermediate between the Bloch vectors of the two quantum states \eqref{eq:states}. The other two POVM elements correspond to Bloch vectors distributed symmetrically along the $z$ axis, in the $x$-$z$ plane of the Bloch sphere. }
	\label{fig:Blochvec}	
\end{figure}

%To get $T(\mathcal{P}_3)$, we take the advantage of the symmetry of the slice $S$. For any $(X,Y)\in T(\calp_k(\delta))$, suppose its preimage is $\boldsymbol{p}\in \calp_k(\delta)$. Then we have $(X,Y)=T((\boldsymbol{p}+\Pi(\boldsymbol{p}))/2)$ due to the definition of $X$ and $Y$. The behavior $(\boldsymbol{p}+\Pi(\boldsymbol{p}))/2$ is invariant to $\Pi$, thus in $\mathcal{P}_\text{sym}(\delta)$. So we have $(X,Y)\in T(\mathcal{P}_\text{sym}(\delta))$, and then $T(\calp_3(\delta))\subseteq T(\mathcal{P}_\text{sym}(\delta))$. Hence, we have $T(\calp_3)= T(\mathcal{P}_\text{sym}(\delta))$. So we only need to look at the boundary of $T(\mathcal{P}_\text{sym}(\delta))$.

To characterize $T(\mathcal{P}_3)$, we take advantage of the symmetry of the slice $S$. Since $\mathcal{P}_k(\delta)$ is closed under $T$, we have $T(\calp_3(\delta))= \mathcal{P}_\text{sym}(\delta)$. Hence we only need to look at the boundary of $\mathcal{P}_\text{sym}(\delta)$. To characterize the boundary of $\mathcal{P}_\text{sym}(\delta)$, again we only need to consider extremal 3-outcome POVMs. 
An extremal 3-outcome POVM, $\{M_0, M_1, M_2\}$, can be parametrized as
\begin{equation} 
	\label{eq:param_bloch}
	M_b=\lambda_b(\mathbb{I}+\boldsymbol{u}_b\cdot\boldsymbol{\sigma}),
\end{equation}
where $|\boldsymbol{u}_b|=1$, $\sum_{b=0}^{2}\lambda_b=1$ and $\sum_{b=0}^{2}\lambda_b\boldsymbol{u}_b=0$. To meet the requirement of normalization, Bloch vectors $\{\boldsymbol{u}_i\}$ of the three-outcome measurement must lie in the same plane. Without loss of generality, we only consider the 3-outcome POVMs that lie in the same plane as the two states (since they represent the effects of all possible measurements on  the states). 
Using the SDP method discussed in Section \ref{sec:witness}, we found that to outline $T(\mathcal{P}_3)$, it is enough to consider extremal 3-outcome POVMs that have a symmetry in the Bloch vectors as depicted in Fig.~\ref{fig:Blochvec}. Indeed, our analytical constructions based on this observation appear to match precisely the results of the SDP methods over 3-outcome POVMs. 
We can thus characterize all extremal 3-outcome POVMs that contribute to the $T(\calp_3(\delta))$ only with one parameter. Here we choose the angle between the $z$-axis and one of the two symmetrical Bloch vector, $\phi\in\left[-\pi/2,\pi/2\right]$. Then we have two Bloch vectors $\boldsymbol{u}_0=(-\sin\phi,0,-\cos\phi)$ and $\boldsymbol{u}_1=(\sin\phi,0,-\cos\phi)$. And we derive $\lambda_0=\lambda_1=1/[2(1+\cos\phi)]$. These symmetric 3-POVMs can be characterized via a single parameter, namely
\begin{align}\label{eq:robustPOVM}
	M_0=&\frac{1}{2(1+\cos\phi)}\begin{pmatrix}
	1-\cos\phi & \sin\phi\\
	\sin\phi & 1+\cos\phi 
\end{pmatrix},\nonumber\\
M_1=&\frac{1}{2(1+\cos\phi)}\begin{pmatrix}
	1-\cos\phi & -\sin\phi\\
	-\sin\phi & 1+\cos\phi 
\end{pmatrix},\nonumber\\
M_2=&\frac{1}{(1+\cos\phi)}\begin{pmatrix}
	2\cos\phi & 0 \\
	0 &0
\end{pmatrix}.
\end{align}
Combined with the states of Eq. \eqref{eq:states}, we get an equation of the boundary in a parametric form:
\begin{align}
	\label{eq:p3curve}
	\begin{cases}
		X=\left[1-\cos(\phi-2\theta)\right]/2(1+\cos\phi),\\
		Y=\left[1-\cos(\phi+2\theta)\right]/2(1+\cos\phi).\\
	\end{cases}
\end{align}

Finally, $T(\calp_3)$ is the convex hull of the trivial point $(0,0)$ and the curve in Eq.~\eqref{eq:p3curve}, as shown in Fig.~\ref{fig:slice_geo}.

Fig.~\ref{fig:slice_delta} shows $T(\mathcal{P}_2(\delta))$ and $T(\mathcal{P}_3(\delta))$ with $\delta=0,0.7,0.9$, and $1$. It again states that the assumption we need to certify whether it is a genuine 3-outcome POVM is merely a lower bound of $\delta$. As $\delta$ varies from $0$ to $1$, $T(\mathcal{P}_2(\delta))$ and $T(\mathcal{P}_3(\delta))$ gradually fills the convex hull of $(0,0)$, $(1,0)$ and $(0,1)$.

	\subsubsection{Robustness against noise}
	\label{subsubsec:robust3-outcome}
	Next we discuss the 3-outcome POVMs that are most robust to white noise, in other words, how much noise we can add to the behavior before it can no longer certify a genuine 3-outcome POVM. This can be investigated using the SDP method above.
	
\begin{figure}[htbp]
		\centering
		\subfloat[]{
		\label{fig:OptOmegaMaxMin}
		\includegraphics[width=2.9in,trim={3cm 9.5cm 3cm 9.6cm},clip]{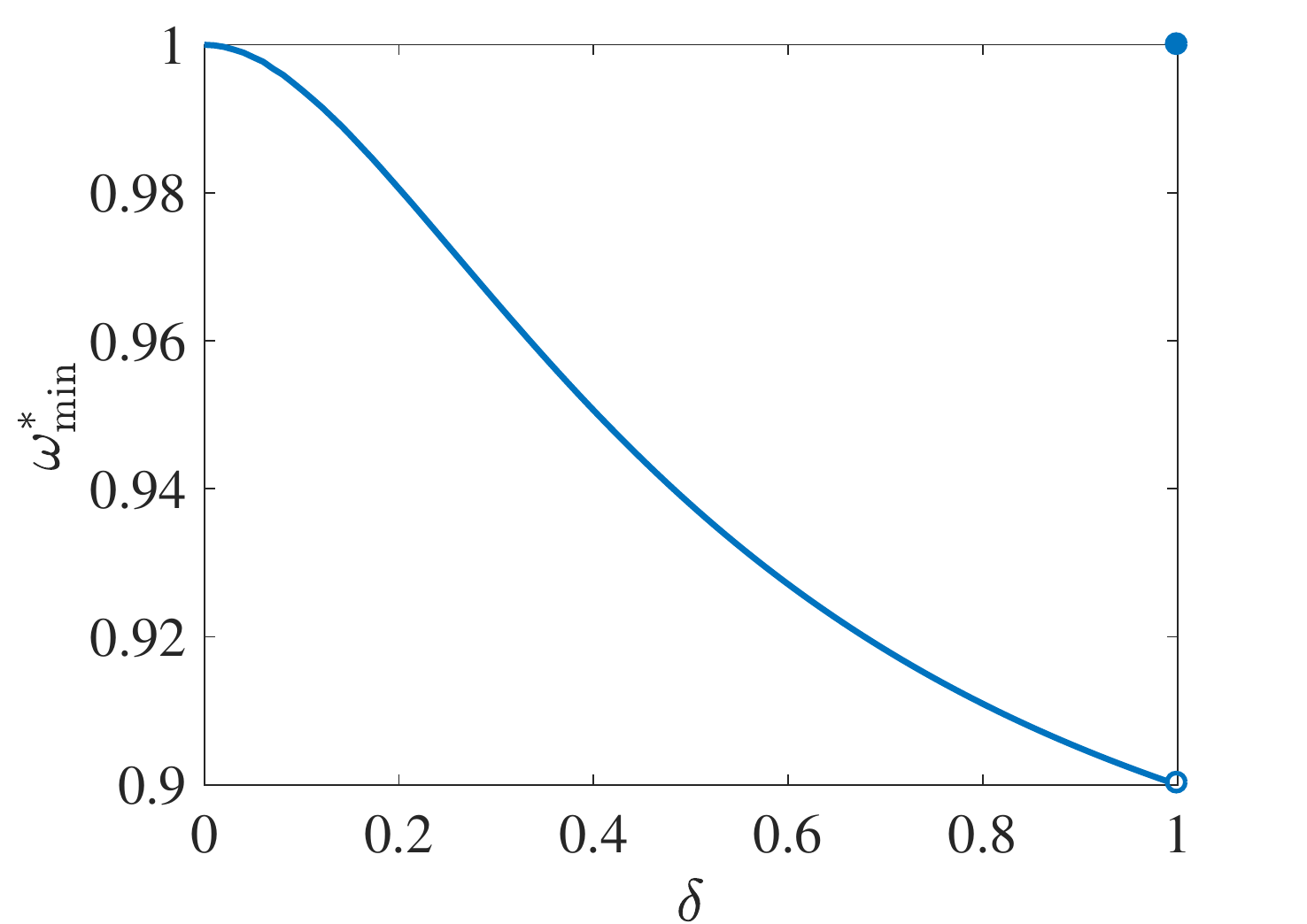}
			}
		
			\subfloat[]{
				\includegraphics[width=3in,trim={3cm 9.5cm 3cm 9.6cm},clip]{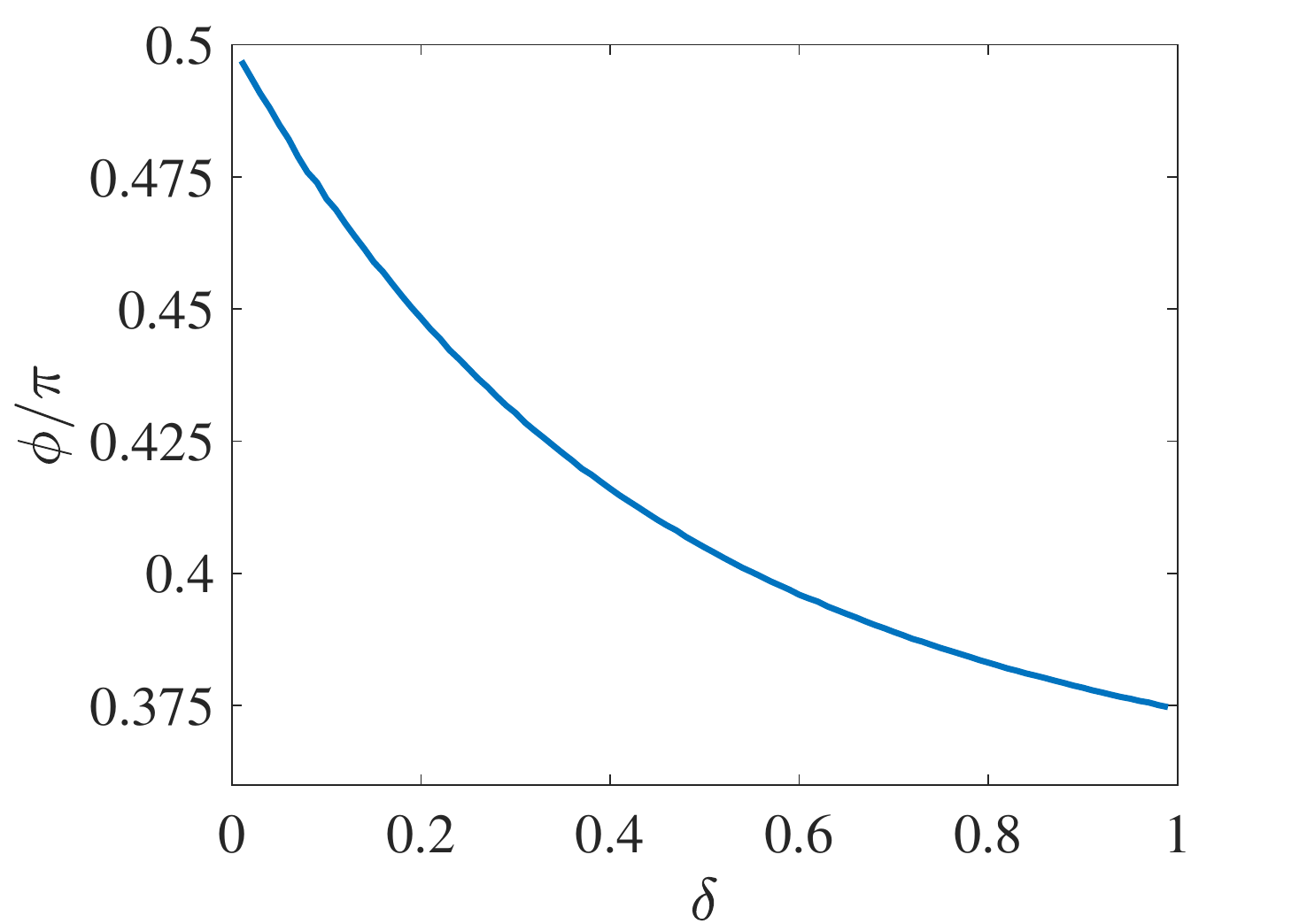}
				\label{fig:phi}
			}
		\caption{
			(a)
			$\omega^*_\text{min}$, which characterizes the robustness of the POVMs against white noise in certification, corresponding to different $\delta$. $\omega^*_\text{min}$ cannot go lower than $0.9$ no matter how close the quantum states are, which means that up to $10\%$ of white noise can be tolerated.
			(b) $\phi$, the only parameter to define a most robust symmetric extremal 3-outcome POVM, as a function of $\delta$.
		}
	\end{figure}

	Here we consider white noise added on the behavior. The robustness against white noise is then characterized by $1-\omega^*$ when we take $\boldsymbol{p}_0=\boldsymbol{p}_\mathbb{I}$.
	Through numerical optimization, we found that the larger the overlap between the quantum states is, the more noise the behavior can tolerate before it falls into $\mathcal{P}_2$.
	In Fig~\ref{fig:OptOmegaMaxMin}, we show the minimal $w^*$ as $\delta$ changes.
	Up to $10\%$ of noise can be tolerated. 
	
	Interestingly, numerical results show that the most robust behavior, i.e. the behavior which gives $\omega^*_\text{min}$, would be on the slice $S$. Hence the corresponding measurement has the symmetric form given in Eq.~\eqref{eq:robustPOVM}, as shown in Fig.~\ref{fig:Blochvec}. For given overlap $\delta$, one can then find numerically the optimal value of $\phi$, characterizing the most robust measurement (see Fig.~\ref{fig:phi}).
	
	The optimal measurement for USD (in Sec.~\ref{sec:usd}) can tolerate at most $4\%$ of white noise, for overlap $\delta=0.46$. For other values of the overlap, the noise tolerance is weaker.
		
%
%Among the POVMs of the form \eqref{eq:robustPOVM}, we searched numerically for the optimal parameter $\phi$ for different value of the overlap $\delta$. Fig.~\ref{fig:figOptOmegaMaxMin} gives $\omega^*_\text{min}$ corresponding to different overlaps, and the optimal $\phi$ is shown in Fig.~\ref{fig:lambdaVsDelta}. Note that since the behavior of POVMs~\eqref{eq:robustPOVM} is in $S$, as plotted in Fig.~\ref{fig:slice_delta}, one can actually solve $\phi$ analytically using our geometrical characterization.

%

\subsection{Unambiguous state discrimination}
\label{sec:usd}
When two states have a non-zero overlap, one cannot perfectly distinguish them. However, if an inconclusive output is allowed in certain instances, this becomes possible via USD~\cite{ivanovic_how_1987,dieks_overlap_1988,peres_how_1988}. Given two states, $\psi_0$ and $\psi_1$, the family of POVMs $\{M_0,M_1,M_\varnothing\}$ that can accomplish the USD task must have $\tr(M_j\ketbra{\psi_{\bar{j}}}{\psi_{\bar{j}}})=0$ due to the unambiguity condition. $M_0$ and $M_1$ are the elements that correspond to the definite answers, and $M_\varnothing$ the inconclusive result. The figure-of-merit in USD is the probability of producing a definite answer, i.e. $p_\mathrm{succ}=(p(0|0)+p(1|1))/2$. If $|\braket{\psi_0|\psi_1}|=\delta$ and the two states have equal occurrence probability, the maximal $p_\text{succ}$ is $1-\delta$ \cite{dieks_overlap_1988,peres_how_1988,ivanovic_how_1987}, denoted by $p_\text{succ,3}$. This requires a genuine 3-outcome POVM (which can be confirmed with the method in Sec.~\ref{sec:analytical}).

\subsubsection{Certifying genuine 3-outcome POVM}
\begin{figure}[htbp]
	\centering
	\includegraphics[width=2.5in,trim={3.2cm 9.5cm 3.2cm 9.5cm},clip]{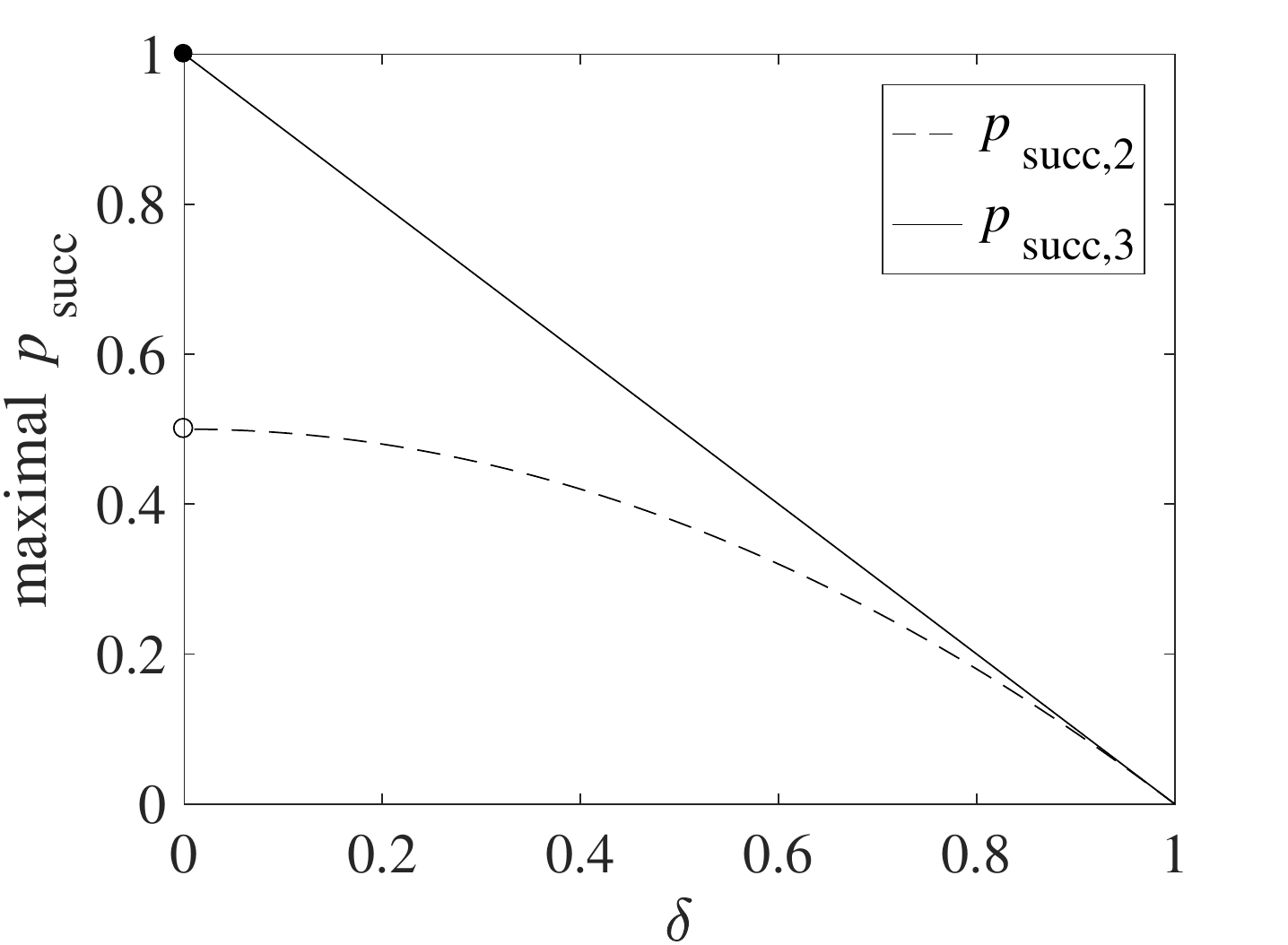}
	\caption{Maximal USD success probability for different lower bounds on overlap. Note the hollow circle at lower bound $0$. In the extreme case where the two input states are orthogonal, they can be perfectly distinguished with a 2-outcome POVM as well. No 2-outcome measurement can distinguish two states unambiguously with success probability larger than $1/2$ when the two states are nonorthogonal ($\delta>0$).}
	\label{fig:p_succ}
\end{figure}

Intuitively, a high $p_\text{succ}$ should certify a genuine 3-outcome POVM. To show this, we upper bound $p_\text{succ}$ restricting ourselves to behaviors in $\calp_2$. To achieve USD, the elements of the POVMs must be orthogonal to the states. To maximize $p_\text{succ}$, it is enough to consider extremal POVMs. Hence the relevant binary POVMs are of the form: $\{\ketbra{\psi_1^\perp}{\psi_1^\perp},0,\mathbb{I}-\ketbra{\psi_1^\perp}{\psi_1^\perp}\}$, and $\{0,\ketbra{\psi_0^\perp}{\psi_0^\perp},\mathbb{I}-\ketbra{\psi_0^\perp}{\psi_0^\perp}\}$, where $\ket{\psi_x^\perp}$ is the orthogonal state of $\ket{\psi_x}$. Due to convexity, one can immediately find that for 2-outcome POVMs, the maximal $p_\mathrm{succ}$ is
\begin{equation}\label{eqn:p_succ_max}
	p_\mathrm{succ,2}=\begin{cases}
	(1-\delta^2)/2, & 0<\delta\leq 1\\
	1, & \delta=0
	\end{cases}.
\end{equation}Since $p_\text{succ,3}>p_\text{succ,2}$ (see Fig.~\ref{fig:p_succ}) when $\delta\in(0,1)$, $p_\text{succ}$ can be used as a witness for genuine 3-outcome POVMs. If the overlap of the states is lower bounded by $\delta$, and $p_\mathrm{succ}$ exceeds Eq.~\eqref{eqn:p_succ_max}, then it certifies a genuine 3-outcome POVM.

Furthermore, one can certify genuine 3-outcome POVMs in terms of $p_\mathrm{succ}$ in a way that is independent from the overlap. For 2-outcome POVMs, when $\delta>0$, $p_\mathrm{succ} \leq1/2$. Thus whenever $p_\mathrm{succ}\ge 1/2$ is observed (and no error occurs), it can be inferred that the measurement box is a genuine 3-outcome POVM.

\subsubsection{Self-testing}
A high success probability for USD not only certifies genuine 3-outcome POVMs, it may even uniquely identify the states and measurement. In this section, we show that under the assumption of bounded overlap $\braket{\psi_0|\psi_1}\geq \delta$, having $p_\text{succ} = 1-\delta$ \textit{self-tests} two qubit states of overlap $\delta$ and the optimal USD measurement. To be precise, following Ref.~\cite{sekatski_certifying_2018}, we say a behavior self-tests the measurement $\{\bar{M}_j\}$ in Hilbert space $\bar{\mathcal{H}}$ if for every quantum realization $(\ket{\psi_0},\ket{\psi_1},\{M_j\})$ in Hilbert space $\mathcal{H}$  compatible with the behavior, there exists a completely positive and trace-preserving (CPTP) map $\Lambda:\mathcal{B}(\bar{\mathcal{H}})\rightarrow\mathcal{B}(\mathcal{H})$, such that  
\begin{equation}
\label{eq:self-test}
	\tr(M_j\Lambda(\ketbra{\bar{\psi}}{\bar{\psi}}))=\tr(\bar{M}_j\ketbra{\bar{\psi}}{\bar{\psi}}),
\end{equation}
is satisfied for any $\ket{\bar{\psi}}\in\bar{\mathcal{H}}$ and $j=0,1,\varnothing$.

In our case, $\bar{\mathcal{H}}=\mathbb{C}^2$, and the ideal states $\ket{\bar{\psi_x}}$ are given in Eq. \eqref{eq:states}, with overlap $\delta$. For the input states, on the one hand we have $|\braket{\psi_0|\psi_1}|\geq \delta$ by assumption, on the other hand $p_\text{succ} = 1-\delta$ implies $|\braket{\psi_0|\psi_1}|\leq \delta$. Hence, $|\braket{\psi_0|\psi_1}|=\delta$.
%because for the behavior yielded by quantum states with $\theta\ge\pi/4$, one can always find a quantum realization with two states with $\theta\leq\pi/4$ by adding a global phase on one of the states, which does not change the result of measuring. Thus, in the following, we will only look at $c^2-s^2\ge0$.}

For the measurements, it is sufficient to construct the map as $\Lambda(\cdot)=K(\cdot)K^\dagger$, where $K: \mathbb{C}^2 \rightarrow \mathcal{H}$ 
\begin{align}
\label{eq:map}
	 \ket{0}&\rightarrow(\ket{\psi_0}+\ket{\psi_1})/2c\nonumber\\ 
	 \ket{1}&\rightarrow(\ket{\psi_0}-\ket{\psi_1})/2s,
\end{align}
and the ideal measurement $\bar{M}$ is:
\begin{align}
    \bar{M}_0&=\frac{1}{1+\delta}\ketbra{\bar{\psi}_1^\perp}{\bar{\psi}_1^\perp},\nonumber\\
    \bar{M}_1&=\frac{1}{1+\delta}\ketbra{\bar{\psi}_0^\perp}{\bar{\psi}_0^\perp},\nonumber\\
    \bar{M}_\varnothing & = \mathbb{I}-\bar{M}_0-\bar{M}_1,
\end{align}
where $\ket{\bar{\psi}_1^\perp}=(s\ket{0}+c\ket{1})$, $\ket{\bar{\psi}_0^\perp}=(s\ket{0}-c\ket{1})$.
It remains to show that Eq.~\eqref{eq:self-test} is satisfied for any qubit states $\rho$. 

Writing an arbitrary qubit state as
$\rho= \sum_{i,j}\rho_{ij}\ketbra{i}{j}$, we have 
\begin{align}
\Lambda(\rho)&=\textstyle\frac{1}{4}\sum_{i,j=0}^{1}\ketbra{\psi_i}{\psi_j} \textstyle\left(\frac{1}{c^2}\rho_{00}+\frac{(-1)^j}{cs}\rho_{01} \right.\nonumber \\
& \left.+\frac{(-1)^i}{cs}\rho_{10}+\frac{(-1)^{i+j}}{s^2}\rho_{11}\right)
\end{align}
From the optimal USD behavior
$$p(b|x)=
\begin{pmatrix}
1-\delta      & 0      &\delta    \\
0      & 1-\delta      & \delta      \\
\end{pmatrix}
$$
(written in the same manner of Eq.~\eqref{eq:distribution_notation}),we have $\tr(M_j\ketbra{\psi_j}{\psi_j})=1-\delta$ and $\tr(M_j\ketbra{\psi_{\bar{j}}}{\psi_{\bar{j}}})=0$.
Exploiting the positivity of $M_k$, we have $\bra{\psi_{\bar{j}}}M_j\ket{\psi_{\bar{j}}}=0 \Leftrightarrow M_j\ket{\psi_{\bar{j}}}=0$, thus
$\tr(M_k\ketbra{\psi_j}{\psi_{j'}})=0$ except 
\begin{align}\label{eq:Ekketbra}
	\tr(M_k\ketbra{\psi_k}{\psi_k})=1-\delta.
\end{align}
Take $\tr(M_0\Lambda(\rho))$ as an example:
\begin{align}
\tr(M_0\Lambda(\rho))=\frac{1}{1+\delta}\left(s^2\rho_{00}+cs\rho_{01}+cs\rho_{10}+c^2\rho_{11}\right).
\end{align}
This is achieved by combining Eq.~\eqref{eq:Ekketbra} and $\delta=c^2-s^2$. By rewriting \[\bar{M}_0=\frac{1}{1+\delta}(s\ket{0}+c\ket{1})(s\bra{0}+c\bra{1}),\] one can arrive at $\tr(\bar{M}_0\rho)=\tr(M_0\Lambda(\rho))$. 
One can check similarly for $M_1$ with $\bar{M}_1$ and $M_\varnothing$ with $\bar{M}_\varnothing$, which completes the proof.

\section{Randomness}
We briefly discuss the connection between our results and the task of randomness generation.
Clearly, the certification of more than one bit of randomness implies a genuine 3-outcome POVM~\cite{ioannou_upper_2019}. 
It turns out however that genuine 3-outcome measurements do not necessarily imply more randomness. There exist genuine 3-outcome POVMs that can certify nearly zero randomness. For example, consider a binary POVM with Bloch vectors aligned with one of the quantum states. From this, we can generate a 3-outcome POVM by slightly rotating and shrinking the POVM elements, and thus allowing a small weight on a third component. In this case, we can obtain a 3-outcome POVM that can be certified to be genuine, but at the same time certifies only little randomness.
As a more concrete example, for $\delta=0.9$, the behavior of the optimal USD measurement can only certify 0.15 bit of randomness (computed via a SDP as in Ref.~\cite{brask_megahertz-rate_2017}).

%the two problems are different, as there exist genuine 3-outcome POVMs that can certify nearly zero randomness. 
%For example, consider a binary POVM with Bloch vectors aligned with one of the quantum states. From this, we can generate a 3-outcome POVM by slightly modifying the POVM elements. In this case, we can obtain a 3-outcome POVM that can be certified to be genuine, but at the same time certifies only little randomness.

\begin{figure}[htbp]
	\centering
	\includegraphics[width=2.8in,trim={3.2cm 9.5cm 3.2cm 9.5cm},clip]{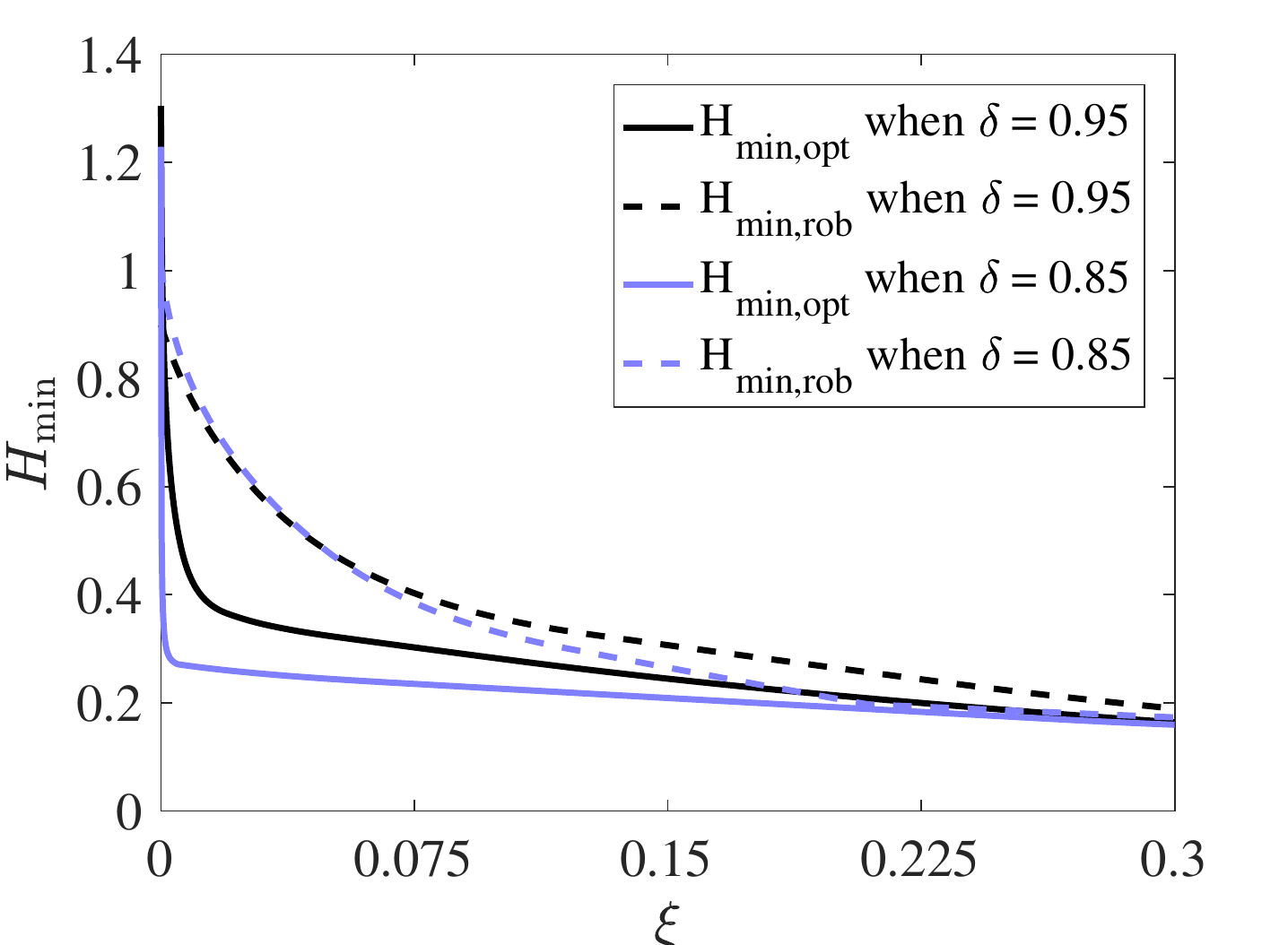}
	\caption{Randomness certifiable by different POVMs mixed with white noise: dashed line for $P_\text{opt}$, and solid line for $P_\text{rob}$. The gap between the randomness is more apparent when $\xi$ is small (e.g. at the given overlaps, when $\xi$ is smaller than $0.2$). When $\xi$ is larger, the two families of measurements give approximatively the same entropy $H_\text{min}$. }
	\label{fig:figRand9}
\end{figure}

Moreover, we investigated the advantage of the optimal POVM for randomness in Ref.~\cite{ioannou_upper_2019}, denoted by $M_\text{opt}$, over other POVMs in the presence of noise. We compare the randomness that can be certified by $M_\text{opt}$ with that of the most robust genuine 3-outcome POVM (discussed in Sec.~\ref{subsubsec:robust3-outcome} and now referred to as $M_\text{rob}$) in the presence of white noise (see Fig.~\ref{fig:figRand9}). That is, for behaviors of the form $\boldsymbol{p}'_\text{opt(rob)}=(1-\xi)\boldsymbol{p}_\text{opt(rob)}+\xi\boldsymbol{p}_\mathbb{I}$, we compute the minimal entropy it can certify as a function of $\xi$. The conclusion is that although $M_\text{opt}$ can certify the most randomness in the ideal, noiseless case, this advantage vanishes once there is noise.

\section{Conclusion}
We discussed the problem of characterizing an unknown POVM in a semi-DI prepare-and-measure scenario, based on the assumption of a minimum overlap between the prepared quantum states. We developed several methods for this problem, and showed how genuine 3-outcome POVM can be certified. 
Furthermore, we showed that it is possible to self-test the optimal measurement for unambiguous state discrimination in this framework.

It would be interesting to see if other properties of quantum systems can be certified in this setting, and if other measurements can be self-tested, in particular in the presence of noise. A relevant problem is the certification of genuine $d$-outcome POVMs, which would require a scenario with at least $d-1$ preparations. In this case, the assumptions of limited distinguishability of the set of prepared states could be formalized in different possible ways.

\section*{Acknowledgments}
We would like to thank Jean-Daniel Bancal, Marie Ioannou, and Davide Rusca for useful discussions. We acknowledge support from the Swiss National Science Foundation (Starting grant DIAQ, Bridge project ``Self-testing QRNG'' and NCCR-QSIT). This work is also supported by funded by National Nature Science Foundation of China (Grants No.~61601476). Weixu Shi is funded by China Scholarship Council.

\appendix

\section{Dual problem} \label{sec:append_dual}
In this section, we show one possible way to dualize the SDP from Eq.~\eqref{eq:primalSDP} to Eq.~\eqref{eq:dualSDP}. We use a similar method as in Ref.~\cite{brask_megahertz-rate_2017}. First we transform the primal problem. Let $N^j_b=\frac{1}{\omega}p_j M^j_b=\frac{1}{\omega}\tilde{M}^j_b$. From the third constraint of Eq.~\eqref{eq:primalSDP} we immediately have $\frac{1}{\omega}=\frac{1}{2}\tr\sum_{j,b}N_b^j$. Since maximizing $\omega$ is equivalent to minimizing $\frac{1}{\omega}$, which we denote by $\eta$, the primal problem can be rewritten as
\begin{align}
    \min_{N^j_b} \quad & \eta=\frac{1}{2}\tr\sum_{j,b}N_b^j\nonumber\\
    \text{subject to} \quad
	& N^j_b \succeq 0, \quad \forall j,b,\nonumber\\
	&\sum_b N^j_b=\frac{1}{2}\tr(\sum_bN^j_b)\mathbb{I}, \quad\forall j,\nonumber\\
	& N^j_j = 0, \quad \forall j,\nonumber\\
	&p(b|x)+(\eta-1)p_0(b|x) \nonumber\\
	& \quad\quad\quad=\tr(\ketbra{\psi_x}{\psi_x}\sum_j N^j_b), \forall x,b.
	\label{eq:primalSDP2}
\end{align}

Introduce Hermitian matrices $G^j_b$, $H^j$, $J^j$, and real scalars $v_{b|x}$ as Lagrange multipliers to each constraints in the primal problem. The Lagrangian associated with Eq.~\eqref{eq:primalSDP2} reads
\begin{align}
	\mathcal{L} = &\frac{1}{2}\sum_{j,b}\tr(N^j_b) + \sum_{j,b}\tr(G^j_bN^j_b)  \nonumber\\
	& +\sum_{j,b}\tr\{H^j[N^j_b-\frac{1}{2}\tr(N^j_b)\mathbb{I}]\} 
	+\sum_{j,b}\delta_{j,b}\tr(N^j_b J^j) \nonumber\\
	& +\sum_{x,b}v_{b|x}\left\lbrace p(b|x)-\tr(\ketbra{\psi_x}{\psi_x}\sum_jN^j_b)\right.\nonumber\\
	& \quad\qquad\left.+p_0(b|x)\left[\frac{1}{2}\sum_{j,b'}\tr(N_{b'}^j)-1\right]\right\rbrace,\label{eq:lagrangian}
\end{align}
where $j,b,b'$ range from $0$ to $2$, and $x$ from $0$ to $1$.

We define $\mathcal{S}$ to be the infimum of the Lagrangian over the primal SDP variables, namely $\mathcal{S}=\inf_{N_b^j}\mathcal{L}$. To let $\mathcal{S}$ be able to lower bound the primal objective function, for any particular solution $N^j_b$, $\mathcal{L}$ should be smaller than the value of the primal problem. In order to achieve this, the second term of Eq.~\eqref{eq:lagrangian} should be negative, which requires $G^j_b\leq0$, while the following three terms vanish automatically for $N^j_b$ that satisfy the constraints in \eqref{eq:primalSDP2}.

Now we maximize $\mathcal{S}$ over the Lagrangian multipliers to get a tighter lower bound of $\mathcal{L}$. By rearranging the terms of Eq.~\eqref{eq:lagrangian}, we have
\begin{equation}
	\mathcal{S}=\sum_{b,x}v_{b|x}\left(p(b|x)-p_0(b|x)\right)+\inf_{N^j_b}\sum_{j,b}\tr\left[N^j_b K^j_b\right]\label{eq:lagrangianS},
\end{equation}
where 
\begin{align}
	K^j_b=&\frac{1}{2}\mathbb{I}+G^j_b+H^j-\frac{1}{2}\tr(H^j)\mathbb{I}+\frac{1}{2}\sum_{x,b'}v_{b'|x}p_0(b'|x)\mathbb{I}\nonumber\\
	&-\sum_x v_{b|x}\ketbra{\psi_x}{\psi_x}+\delta_{j,b}J^j \label{eq:lagrangianK}
\end{align}
Since there is no constraint on $N^j_b$ in the Lagrangian, to make Eq.~\eqref{eq:lagrangianS} nontrivial, namely, $\mathcal{S}>-\infty$, $K^j_b$ is restricted to be zero. We can solve $K^j_b=0$ for $G^j_b$ and substitute it into $G^j_b\leq0$, which is the third constraint of Eq.~\eqref{eq:dualSDP}.

\section{Detailed calculations for the analytic boundary of $\calp_2(\delta)$ in the symmetric slice} 
\label{sec:append_Characterzation of slice}

To characterize $T(\calp_2(\delta))$, we write the two quantum states as $\{\frac{1}{2}(\mathbb{I}+\boldsymbol{n}_x\cdot\boldsymbol{\sigma})\}_{x=0}^1$, and the projective 2-outcome POVMs as $\{\frac{1}{2}(\mathbb{I}\pm\boldsymbol{u}\cdot\boldsymbol{\sigma})\}$, where $\boldsymbol{n}_x$ and $\boldsymbol{u}$ are the Bloch vectors, and $\boldsymbol{\sigma}$ is the vector of Pauli operators. According to Eq.~\eqref{eq:states}, $n_x=((-1)^x\sin2\theta,0,\cos2\theta)$. For strategy $\{0,K_1,\mathbb{I}-K_1\}$ we have
\begin{align}
	X=\tfrac{1}{4}\left(1+\boldsymbol{n_1}\cdot\boldsymbol{u}\right)\nonumber\\
	Y=\tfrac{1}{4}\left(1+\boldsymbol{n_0}\cdot\boldsymbol{u}\right).
\end{align}
We find that 
\begin{align}
\label{eq:XY}
	X+Y&=\frac{1}{2}+\frac{1}{4}(\boldsymbol{n}_0+\boldsymbol{n}_1)\cdot\boldsymbol{u}\nonumber\\
		X-Y&=\frac{1}{4}(\boldsymbol{n}_0-\boldsymbol{n}_1)\cdot\boldsymbol{u}
\end{align}
Since $(\boldsymbol{n_0}+\boldsymbol{n_1})\perp(\boldsymbol{n_0}-\boldsymbol{n_1})$ and $\boldsymbol{u}$ is a unit vector, we have 
\begin{align}
\label{eq:cossin}
	\left[\frac{\boldsymbol{u}\cdot(\boldsymbol{n_0}+\boldsymbol{n_1})}{|\boldsymbol{n_0}+\boldsymbol{n_1}|}\right]^2+\left[\frac{\boldsymbol{u}\cdot(\boldsymbol{n_0}-\boldsymbol{n_1})}{|\boldsymbol{n_0}-\boldsymbol{n_1}|}\right]^2=1.
\end{align}
Rewriting Eq.~\eqref{eq:cossin} in terms of Eq.~\eqref{eq:XY} leads to Eq.~\eqref{eq:P2ellipse}.

This works for strategy $\{K_2,0,\mathbb{I}-K_2\}$ also. As to strategy $\{K_3,\mathbb{I}-K_3,0\}$, immediately we have $X+Y=1$, but not all the points on the line are accessible. Note that $$X/Y = \frac{1+\frac{1}{2}\boldsymbol{u}_0\cdot(\boldsymbol{n}_0-\boldsymbol{n}_1)}{1-\frac{1}{2}\boldsymbol{u}_0\cdot(\boldsymbol{n}_0-\boldsymbol{n}_1)}\in\left[\tfrac{1-\sqrt{1-\delta^2}}{1+\sqrt{1-\delta^2}},\tfrac{1+\sqrt{1-\delta^2}}{1-\sqrt{1-\delta^2}}\right],$$
we have that only the line segment between vertices \eqref{eq:vertice1} and \eqref{eq:vertice2} is valid. Combined with the vertices contributed by trivial measurements, we know that the $(X,Y)$ allowed by the convex combination of 2-outcome POVMs is the convex hull of points $\{(0,0),\text{Eq.~}\eqref{eq:vertice1},\text{Eq.~}\eqref{eq:vertice2}\}$ and the ellipse~\eqref{eq:P2ellipse}.

\bibliography{reference}

\begin{thebibliography}{39}%
\makeatletter
\providecommand \@ifxundefined [1]{%
 \@ifx{#1\undefined}
}%
\providecommand \@ifnum [1]{%
 \ifnum #1\expandafter \@firstoftwo
 \else \expandafter \@secondoftwo
 \fi
}%
\providecommand \@ifx [1]{%
 \ifx #1\expandafter \@firstoftwo
 \else \expandafter \@secondoftwo
 \fi
}%
\providecommand \natexlab [1]{#1}%
\providecommand \enquote  [1]{``#1''}%
\providecommand \bibnamefont  [1]{#1}%
\providecommand \bibfnamefont [1]{#1}%
\providecommand \citenamefont [1]{#1}%
\providecommand \href@noop [0]{\@secondoftwo}%
\providecommand \href [0]{\begingroup \@sanitize@url \@href}%
\providecommand \@href[1]{\@@startlink{#1}\@@href}%
\providecommand \@@href[1]{\endgroup#1\@@endlink}%
\providecommand \@sanitize@url [0]{\catcode `\\12\catcode `\$12\catcode
  `\&12\catcode `\#12\catcode `\^12\catcode `\_12\catcode `\%12\relax}%
\providecommand \@@startlink[1]{}%
\providecommand \@@endlink[0]{}%
\providecommand \url  [0]{\begingroup\@sanitize@url \@url }%
\providecommand \@url [1]{\endgroup\@href {#1}{\urlprefix }}%
\providecommand \urlprefix  [0]{URL }%
\providecommand \Eprint [0]{\href }%
\providecommand \doibase [0]{http://dx.doi.org/}%
\providecommand \selectlanguage [0]{\@gobble}%
\providecommand \bibinfo  [0]{\@secondoftwo}%
\providecommand \bibfield  [0]{\@secondoftwo}%
\providecommand \translation [1]{[#1]}%
\providecommand \BibitemOpen [0]{}%
\providecommand \bibitemStop [0]{}%
\providecommand \bibitemNoStop [0]{.\EOS\space}%
\providecommand \EOS [0]{\spacefactor3000\relax}%
\providecommand \BibitemShut  [1]{\csname bibitem#1\endcsname}%
\let\auto@bib@innerbib\@empty
%</preamble>
\bibitem [{\citenamefont {Barrett}\ \emph {et~al.}(2005)\citenamefont
  {Barrett}, \citenamefont {Hardy},\ and\ \citenamefont
  {Kent}}]{barrett_no_2005}%
  \BibitemOpen
  \bibfield  {author} {\bibinfo {author} {\bibfnamefont {J.}~\bibnamefont
  {Barrett}}, \bibinfo {author} {\bibfnamefont {L.}~\bibnamefont {Hardy}}, \
  and\ \bibinfo {author} {\bibfnamefont {A.}~\bibnamefont {Kent}},\ }\href
  {\doibase 10.1103/PhysRevLett.95.010503} {\bibfield  {journal} {\bibinfo
  {journal} {Phys. Rev. Lett.}\ }\textbf {\bibinfo {volume} {95}},\ \bibinfo
  {pages} {010503} (\bibinfo {year} {2005})}\BibitemShut {NoStop}%
\bibitem [{\citenamefont {Acín}\ \emph {et~al.}(2007)\citenamefont {Acín},
  \citenamefont {Brunner}, \citenamefont {Gisin}, \citenamefont {Massar},
  \citenamefont {Pironio},\ and\ \citenamefont
  {Scarani}}]{acin_device-independent_2007}%
  \BibitemOpen
  \bibfield  {author} {\bibinfo {author} {\bibfnamefont {A.}~\bibnamefont
  {Acín}}, \bibinfo {author} {\bibfnamefont {N.}~\bibnamefont {Brunner}},
  \bibinfo {author} {\bibfnamefont {N.}~\bibnamefont {Gisin}}, \bibinfo
  {author} {\bibfnamefont {S.}~\bibnamefont {Massar}}, \bibinfo {author}
  {\bibfnamefont {S.}~\bibnamefont {Pironio}}, \ and\ \bibinfo {author}
  {\bibfnamefont {V.}~\bibnamefont {Scarani}},\ }\href {\doibase
  10.1103/PhysRevLett.98.230501} {\bibfield  {journal} {\bibinfo  {journal}
  {Phys. Rev. Lett.}\ }\textbf {\bibinfo {volume} {98}},\ \bibinfo {pages}
  {230501} (\bibinfo {year} {2007})}\BibitemShut {NoStop}%
\bibitem [{\citenamefont {Colbeck}(2009)}]{colbeck_quantum_2009}%
  \BibitemOpen
  \bibfield  {author} {\bibinfo {author} {\bibfnamefont {R.}~\bibnamefont
  {Colbeck}},\ }\href {http://arxiv.org/abs/0911.3814} {\bibfield  {journal}
  {\bibinfo  {journal} {arXiv:0911.3814 [quant-ph]}\ } (\bibinfo {year}
  {2009})}\BibitemShut {NoStop}%
\bibitem [{\citenamefont {Pironio}\ \emph {et~al.}(2010)\citenamefont
  {Pironio}, \citenamefont {Acín}, \citenamefont {Massar}, \citenamefont
  {de~la Giroday}, \citenamefont {Matsukevich}, \citenamefont {Maunz},
  \citenamefont {Olmschenk}, \citenamefont {Hayes}, \citenamefont {Luo},
  \citenamefont {Manning},\ and\ \citenamefont {Monroe}}]{pironio_random_2010}%
  \BibitemOpen
  \bibfield  {author} {\bibinfo {author} {\bibfnamefont {S.}~\bibnamefont
  {Pironio}}, \bibinfo {author} {\bibfnamefont {A.}~\bibnamefont {Acín}},
  \bibinfo {author} {\bibfnamefont {S.}~\bibnamefont {Massar}}, \bibinfo
  {author} {\bibfnamefont {A.~B.}\ \bibnamefont {de~la Giroday}}, \bibinfo
  {author} {\bibfnamefont {D.~N.}\ \bibnamefont {Matsukevich}}, \bibinfo
  {author} {\bibfnamefont {P.}~\bibnamefont {Maunz}}, \bibinfo {author}
  {\bibfnamefont {S.}~\bibnamefont {Olmschenk}}, \bibinfo {author}
  {\bibfnamefont {D.}~\bibnamefont {Hayes}}, \bibinfo {author} {\bibfnamefont
  {L.}~\bibnamefont {Luo}}, \bibinfo {author} {\bibfnamefont {T.~A.}\
  \bibnamefont {Manning}}, \ and\ \bibinfo {author} {\bibfnamefont
  {C.}~\bibnamefont {Monroe}},\ }\href {\doibase 10.1038/nature09008}
  {\bibfield  {journal} {\bibinfo  {journal} {Nature}\ }\textbf {\bibinfo
  {volume} {464}},\ \bibinfo {pages} {1021} (\bibinfo {year}
  {2010})}\BibitemShut {NoStop}%
\bibitem [{\citenamefont {Scarani}\ and\ \citenamefont
  {Kurtsiefer}(2014)}]{scarani_black_2014}%
  \BibitemOpen
  \bibfield  {author} {\bibinfo {author} {\bibfnamefont {V.}~\bibnamefont
  {Scarani}}\ and\ \bibinfo {author} {\bibfnamefont {C.}~\bibnamefont
  {Kurtsiefer}},\ }\href {\doibase 10.1016/j.tcs.2014.09.015} {\bibfield
  {journal} {\bibinfo  {journal} {Theoretical Computer Science}\ }\bibinfo
  {series} {Theoretical {Aspects} of {Quantum} {Cryptography} – celebrating
  30 years of {BB}84},\ \textbf {\bibinfo {volume} {560}},\ \bibinfo {pages}
  {27} (\bibinfo {year} {2014})}\BibitemShut {NoStop}%
\bibitem [{\citenamefont {Brunner}\ \emph {et~al.}(2014)\citenamefont
  {Brunner}, \citenamefont {Cavalcanti}, \citenamefont {Pironio}, \citenamefont
  {Scarani},\ and\ \citenamefont {Wehner}}]{brunner_bell_2014}%
  \BibitemOpen
  \bibfield  {author} {\bibinfo {author} {\bibfnamefont {N.}~\bibnamefont
  {Brunner}}, \bibinfo {author} {\bibfnamefont {D.}~\bibnamefont {Cavalcanti}},
  \bibinfo {author} {\bibfnamefont {S.}~\bibnamefont {Pironio}}, \bibinfo
  {author} {\bibfnamefont {V.}~\bibnamefont {Scarani}}, \ and\ \bibinfo
  {author} {\bibfnamefont {S.}~\bibnamefont {Wehner}},\ }\href {\doibase
  10.1103/RevModPhys.86.419} {\bibfield  {journal} {\bibinfo  {journal} {Rev.
  Mod. Phys.}\ }\textbf {\bibinfo {volume} {86}},\ \bibinfo {pages} {419}
  (\bibinfo {year} {2014})}\BibitemShut {NoStop}%
\bibitem [{\citenamefont {Mayers}\ and\ \citenamefont
  {Yao}(2004)}]{mayers_self_2003}%
  \BibitemOpen
  \bibfield  {author} {\bibinfo {author} {\bibfnamefont {D.}~\bibnamefont
  {Mayers}}\ and\ \bibinfo {author} {\bibfnamefont {A.}~\bibnamefont {Yao}},\
  }\href {http://dl.acm.org/citation.cfm?id=2011827.2011830} {\bibfield
  {journal} {\bibinfo  {journal} {Quantum Info. Comput.}\ }\textbf {\bibinfo
  {volume} {4}},\ \bibinfo {pages} {273} (\bibinfo {year} {2004})}\BibitemShut
  {NoStop}%
\bibitem [{\citenamefont {Summers}\ and\ \citenamefont
  {Werner}(1987)}]{summers_maximal_1987}%
  \BibitemOpen
  \bibfield  {author} {\bibinfo {author} {\bibfnamefont {S.~J.}\ \bibnamefont
  {Summers}}\ and\ \bibinfo {author} {\bibfnamefont {R.}~\bibnamefont
  {Werner}},\ }\href {https://projecteuclid.org/euclid.cmp/1104159237}
  {\bibfield  {journal} {\bibinfo  {journal} {Comm. Math. Phys.}\ }\textbf
  {\bibinfo {volume} {110}},\ \bibinfo {pages} {247} (\bibinfo {year}
  {1987})}\BibitemShut {NoStop}%
\bibitem [{\citenamefont {Popescu}\ and\ \citenamefont
  {Rohrlich}(1992)}]{popescu_generic_1992}%
  \BibitemOpen
  \bibfield  {author} {\bibinfo {author} {\bibfnamefont {S.}~\bibnamefont
  {Popescu}}\ and\ \bibinfo {author} {\bibfnamefont {D.}~\bibnamefont
  {Rohrlich}},\ }\href {\doibase 10.1016/0375-9601(92)90711-T} {\bibfield
  {journal} {\bibinfo  {journal} {Phys. Lett. A}\ }\textbf {\bibinfo {volume}
  {166}},\ \bibinfo {pages} {293} (\bibinfo {year} {1992})}\BibitemShut
  {NoStop}%
\bibitem [{\citenamefont {McKague}\ \emph {et~al.}(2012)\citenamefont
  {McKague}, \citenamefont {Yang},\ and\ \citenamefont
  {Scarani}}]{mckague_robust_2012}%
  \BibitemOpen
  \bibfield  {author} {\bibinfo {author} {\bibfnamefont {M.}~\bibnamefont
  {McKague}}, \bibinfo {author} {\bibfnamefont {T.~H.}\ \bibnamefont {Yang}}, \
  and\ \bibinfo {author} {\bibfnamefont {V.}~\bibnamefont {Scarani}},\ }\href
  {\doibase 10.1088/1751-8113/45/45/455304} {\bibfield  {journal} {\bibinfo
  {journal} {J. Phys. A: Math. Theor.}\ }\textbf {\bibinfo {volume} {45}},\
  \bibinfo {pages} {455304} (\bibinfo {year} {2012})}\BibitemShut {NoStop}%
\bibitem [{\citenamefont {Kaniewski}(2016)}]{kaniewski_analytic_2016}%
  \BibitemOpen
  \bibfield  {author} {\bibinfo {author} {\bibfnamefont {J.}~\bibnamefont
  {Kaniewski}},\ }\href {\doibase 10.1103/PhysRevLett.117.070402} {\bibfield
  {journal} {\bibinfo  {journal} {Phys. Rev. Lett.}\ }\textbf {\bibinfo
  {volume} {117}},\ \bibinfo {pages} {070402} (\bibinfo {year}
  {2016})}\BibitemShut {NoStop}%
\bibitem [{\citenamefont {Gallego}\ \emph {et~al.}(2010)\citenamefont
  {Gallego}, \citenamefont {Brunner}, \citenamefont {Hadley},\ and\
  \citenamefont {Acín}}]{gallego_device-independent_2010}%
  \BibitemOpen
  \bibfield  {author} {\bibinfo {author} {\bibfnamefont {R.}~\bibnamefont
  {Gallego}}, \bibinfo {author} {\bibfnamefont {N.}~\bibnamefont {Brunner}},
  \bibinfo {author} {\bibfnamefont {C.}~\bibnamefont {Hadley}}, \ and\ \bibinfo
  {author} {\bibfnamefont {A.}~\bibnamefont {Acín}},\ }\href {\doibase
  10.1103/PhysRevLett.105.230501} {\bibfield  {journal} {\bibinfo  {journal}
  {Phys. Rev. Lett.}\ }\textbf {\bibinfo {volume} {105}},\ \bibinfo {pages}
  {230501} (\bibinfo {year} {2010})}\BibitemShut {NoStop}%
\bibitem [{\citenamefont {Wehner}\ \emph {et~al.}(2008)\citenamefont {Wehner},
  \citenamefont {Christandl},\ and\ \citenamefont
  {Doherty}}]{wehner_lower_2008}%
  \BibitemOpen
  \bibfield  {author} {\bibinfo {author} {\bibfnamefont {S.}~\bibnamefont
  {Wehner}}, \bibinfo {author} {\bibfnamefont {M.}~\bibnamefont {Christandl}},
  \ and\ \bibinfo {author} {\bibfnamefont {A.~C.}\ \bibnamefont {Doherty}},\
  }\href {\doibase 10.1103/PhysRevA.78.062112} {\bibfield  {journal} {\bibinfo
  {journal} {Phys. Rev. A}\ }\textbf {\bibinfo {volume} {78}},\ \bibinfo
  {pages} {062112} (\bibinfo {year} {2008})}\BibitemShut {NoStop}%
\bibitem [{\citenamefont {Bowles}\ \emph {et~al.}(2014)\citenamefont {Bowles},
  \citenamefont {Quintino},\ and\ \citenamefont
  {Brunner}}]{bowles_certifying_2014}%
  \BibitemOpen
  \bibfield  {author} {\bibinfo {author} {\bibfnamefont {J.}~\bibnamefont
  {Bowles}}, \bibinfo {author} {\bibfnamefont {M.~T.}\ \bibnamefont
  {Quintino}}, \ and\ \bibinfo {author} {\bibfnamefont {N.}~\bibnamefont
  {Brunner}},\ }\href {\doibase 10.1103/PhysRevLett.112.140407} {\bibfield
  {journal} {\bibinfo  {journal} {Phys. Rev. Lett.}\ }\textbf {\bibinfo
  {volume} {112}},\ \bibinfo {pages} {140407} (\bibinfo {year}
  {2014})}\BibitemShut {NoStop}%
\bibitem [{\citenamefont {Sekatski}\ \emph {et~al.}(2018)\citenamefont
  {Sekatski}, \citenamefont {Bancal}, \citenamefont {Wagner},\ and\
  \citenamefont {Sangouard}}]{sekatski_certifying_2018}%
  \BibitemOpen
  \bibfield  {author} {\bibinfo {author} {\bibfnamefont {P.}~\bibnamefont
  {Sekatski}}, \bibinfo {author} {\bibfnamefont {J.-D.}\ \bibnamefont
  {Bancal}}, \bibinfo {author} {\bibfnamefont {S.}~\bibnamefont {Wagner}}, \
  and\ \bibinfo {author} {\bibfnamefont {N.}~\bibnamefont {Sangouard}},\ }\href
  {\doibase 10.1103/PhysRevLett.121.180505} {\bibfield  {journal} {\bibinfo
  {journal} {Phys. Rev. Lett.}\ }\textbf {\bibinfo {volume} {121}},\ \bibinfo
  {pages} {180505} (\bibinfo {year} {2018})}\BibitemShut {NoStop}%
\bibitem [{\citenamefont {Tavakoli}\ \emph
  {et~al.}(2018{\natexlab{a}})\citenamefont {Tavakoli}, \citenamefont
  {Kaniewski}, \citenamefont {Vértesi}, \citenamefont {Rosset},\ and\
  \citenamefont {Brunner}}]{tavakoli_self-testing_2018}%
  \BibitemOpen
  \bibfield  {author} {\bibinfo {author} {\bibfnamefont {A.}~\bibnamefont
  {Tavakoli}}, \bibinfo {author} {\bibfnamefont {J.}~\bibnamefont {Kaniewski}},
  \bibinfo {author} {\bibfnamefont {T.}~\bibnamefont {Vértesi}}, \bibinfo
  {author} {\bibfnamefont {D.}~\bibnamefont {Rosset}}, \ and\ \bibinfo {author}
  {\bibfnamefont {N.}~\bibnamefont {Brunner}},\ }\href {\doibase
  10.1103/PhysRevA.98.062307} {\bibfield  {journal} {\bibinfo  {journal} {Phys.
  Rev. A}\ }\textbf {\bibinfo {volume} {98}},\ \bibinfo {pages} {062307}
  (\bibinfo {year} {2018}{\natexlab{a}})}\BibitemShut {NoStop}%
\bibitem [{\citenamefont {Farkas}\ and\ \citenamefont
  {Kaniewski}(2019)}]{farkas_self-testing_2019}%
  \BibitemOpen
  \bibfield  {author} {\bibinfo {author} {\bibfnamefont {M.}~\bibnamefont
  {Farkas}}\ and\ \bibinfo {author} {\bibfnamefont {J.}~\bibnamefont
  {Kaniewski}},\ }\href {\doibase 10.1103/PhysRevA.99.032316} {\bibfield
  {journal} {\bibinfo  {journal} {Phys. Rev. A}\ }\textbf {\bibinfo {volume}
  {99}},\ \bibinfo {pages} {032316} (\bibinfo {year} {2019})}\BibitemShut
  {NoStop}%
\bibitem [{\citenamefont {Tavakoli}\ \emph {et~al.}(2019)\citenamefont
  {Tavakoli}, \citenamefont {Rosset},\ and\ \citenamefont
  {Renou}}]{tavakoli_enabling_2019}%
  \BibitemOpen
  \bibfield  {author} {\bibinfo {author} {\bibfnamefont {A.}~\bibnamefont
  {Tavakoli}}, \bibinfo {author} {\bibfnamefont {D.}~\bibnamefont {Rosset}}, \
  and\ \bibinfo {author} {\bibfnamefont {M.-O.}\ \bibnamefont {Renou}},\ }\href
  {\doibase 10.1103/PhysRevLett.122.070501} {\bibfield  {journal} {\bibinfo
  {journal} {Phys. Rev. Lett.}\ }\textbf {\bibinfo {volume} {122}},\ \bibinfo
  {pages} {070501} (\bibinfo {year} {2019})}\BibitemShut {NoStop}%
\bibitem [{\citenamefont {Tavakoli}\ \emph
  {et~al.}(2018{\natexlab{b}})\citenamefont {Tavakoli}, \citenamefont {Smania},
  \citenamefont {Vértesi}, \citenamefont {Brunner},\ and\ \citenamefont
  {Bourennane}}]{tavakoli_self-testing_2018-1}%
  \BibitemOpen
  \bibfield  {author} {\bibinfo {author} {\bibfnamefont {A.}~\bibnamefont
  {Tavakoli}}, \bibinfo {author} {\bibfnamefont {M.}~\bibnamefont {Smania}},
  \bibinfo {author} {\bibfnamefont {T.}~\bibnamefont {Vértesi}}, \bibinfo
  {author} {\bibfnamefont {N.}~\bibnamefont {Brunner}}, \ and\ \bibinfo
  {author} {\bibfnamefont {M.}~\bibnamefont {Bourennane}},\ }\href
  {http://arxiv.org/abs/1811.12712} {\bibfield  {journal} {\bibinfo  {journal}
  {arXiv:1811.12712 [quant-ph]}\ } (\bibinfo {year}
  {2018}{\natexlab{b}})}\BibitemShut {NoStop}%
\bibitem [{\citenamefont {Miklin}\ \emph {et~al.}(2019)\citenamefont {Miklin},
  \citenamefont {Borkała},\ and\ \citenamefont
  {Pawłowski}}]{miklin_self-testing_2019}%
  \BibitemOpen
  \bibfield  {author} {\bibinfo {author} {\bibfnamefont {N.}~\bibnamefont
  {Miklin}}, \bibinfo {author} {\bibfnamefont {J.~J.}\ \bibnamefont
  {Borkała}}, \ and\ \bibinfo {author} {\bibfnamefont {M.}~\bibnamefont
  {Pawłowski}},\ }\href {http://arxiv.org/abs/1903.12533} {\bibfield
  {journal} {\bibinfo  {journal} {arXiv:1903.12533 [quant-ph]}\ } (\bibinfo
  {year} {2019})}\BibitemShut {NoStop}%
\bibitem [{\citenamefont {Mironowicz}\ and\ \citenamefont
  {Pawłowski}(2018)}]{mironowicz_experimentally_2018}%
  \BibitemOpen
  \bibfield  {author} {\bibinfo {author} {\bibfnamefont {P.}~\bibnamefont
  {Mironowicz}}\ and\ \bibinfo {author} {\bibfnamefont {M.}~\bibnamefont
  {Pawłowski}},\ }\href {http://arxiv.org/abs/1811.12872} {\bibfield
  {journal} {\bibinfo  {journal} {arXiv:1811.12872 [quant-ph]}\ } (\bibinfo
  {year} {2018})}\BibitemShut {NoStop}%
\bibitem [{\citenamefont {Mohan}\ \emph {et~al.}(2019)\citenamefont {Mohan},
  \citenamefont {Tavakoli},\ and\ \citenamefont {Brunner}}]{Mohan_2019}%
  \BibitemOpen
  \bibfield  {author} {\bibinfo {author} {\bibfnamefont {K.}~\bibnamefont
  {Mohan}}, \bibinfo {author} {\bibfnamefont {A.}~\bibnamefont {Tavakoli}}, \
  and\ \bibinfo {author} {\bibfnamefont {N.}~\bibnamefont {Brunner}},\ }\href
  {\doibase 10.1088/1367-2630/ab3773} {\bibfield  {journal} {\bibinfo
  {journal} {New Journal of Physics}\ }\textbf {\bibinfo {volume} {21}},\
  \bibinfo {pages} {083034} (\bibinfo {year} {2019})}\BibitemShut {NoStop}%
\bibitem [{\citenamefont {Pawłowski}\ and\ \citenamefont
  {Brunner}(2011)}]{pawlowski_semi-device-independent_2011}%
  \BibitemOpen
  \bibfield  {author} {\bibinfo {author} {\bibfnamefont {M.}~\bibnamefont
  {Pawłowski}}\ and\ \bibinfo {author} {\bibfnamefont {N.}~\bibnamefont
  {Brunner}},\ }\href {\doibase 10.1103/PhysRevA.84.010302} {\bibfield
  {journal} {\bibinfo  {journal} {Phys. Rev. A}\ }\textbf {\bibinfo {volume}
  {84}},\ \bibinfo {pages} {010302} (\bibinfo {year} {2011})}\BibitemShut
  {NoStop}%
\bibitem [{\citenamefont {Li}\ \emph {et~al.}(2012)\citenamefont {Li},
  \citenamefont {Pawłowski}, \citenamefont {Yin}, \citenamefont {Guo},\ and\
  \citenamefont {Han}}]{li_semi-device-independent_2012}%
  \BibitemOpen
  \bibfield  {author} {\bibinfo {author} {\bibfnamefont {H.-W.}\ \bibnamefont
  {Li}}, \bibinfo {author} {\bibfnamefont {M.}~\bibnamefont {Pawłowski}},
  \bibinfo {author} {\bibfnamefont {Z.-Q.}\ \bibnamefont {Yin}}, \bibinfo
  {author} {\bibfnamefont {G.-C.}\ \bibnamefont {Guo}}, \ and\ \bibinfo
  {author} {\bibfnamefont {Z.-F.}\ \bibnamefont {Han}},\ }\href {\doibase
  10.1103/PhysRevA.85.052308} {\bibfield  {journal} {\bibinfo  {journal} {Phys.
  Rev. A}\ }\textbf {\bibinfo {volume} {85}},\ \bibinfo {pages} {052308}
  (\bibinfo {year} {2012})}\BibitemShut {NoStop}%
\bibitem [{\citenamefont {Lunghi}\ \emph {et~al.}(2015)\citenamefont {Lunghi},
  \citenamefont {Brask}, \citenamefont {Lim}, \citenamefont {Lavigne},
  \citenamefont {Bowles}, \citenamefont {Martin}, \citenamefont {Zbinden},\
  and\ \citenamefont {Brunner}}]{lunghi_self-testing_2015}%
  \BibitemOpen
  \bibfield  {author} {\bibinfo {author} {\bibfnamefont {T.}~\bibnamefont
  {Lunghi}}, \bibinfo {author} {\bibfnamefont {J.~B.}\ \bibnamefont {Brask}},
  \bibinfo {author} {\bibfnamefont {C.~C.~W.}\ \bibnamefont {Lim}}, \bibinfo
  {author} {\bibfnamefont {Q.}~\bibnamefont {Lavigne}}, \bibinfo {author}
  {\bibfnamefont {J.}~\bibnamefont {Bowles}}, \bibinfo {author} {\bibfnamefont
  {A.}~\bibnamefont {Martin}}, \bibinfo {author} {\bibfnamefont
  {H.}~\bibnamefont {Zbinden}}, \ and\ \bibinfo {author} {\bibfnamefont
  {N.}~\bibnamefont {Brunner}},\ }\href {\doibase
  10.1103/PhysRevLett.114.150501} {\bibfield  {journal} {\bibinfo  {journal}
  {Phys. Rev. Lett.}\ }\textbf {\bibinfo {volume} {114}},\ \bibinfo {pages}
  {150501} (\bibinfo {year} {2015})}\BibitemShut {NoStop}%
\bibitem [{\citenamefont {Bourennane}\ \emph {et~al.}(2004)\citenamefont
  {Bourennane}, \citenamefont {Eibl}, \citenamefont {Kurtsiefer}, \citenamefont
  {Gaertner}, \citenamefont {Weinfurter}, \citenamefont {Gühne}, \citenamefont
  {Hyllus}, \citenamefont {Bruß}, \citenamefont {Lewenstein},\ and\
  \citenamefont {Sanpera}}]{bourennane_experimental_2004}%
  \BibitemOpen
  \bibfield  {author} {\bibinfo {author} {\bibfnamefont {M.}~\bibnamefont
  {Bourennane}}, \bibinfo {author} {\bibfnamefont {M.}~\bibnamefont {Eibl}},
  \bibinfo {author} {\bibfnamefont {C.}~\bibnamefont {Kurtsiefer}}, \bibinfo
  {author} {\bibfnamefont {S.}~\bibnamefont {Gaertner}}, \bibinfo {author}
  {\bibfnamefont {H.}~\bibnamefont {Weinfurter}}, \bibinfo {author}
  {\bibfnamefont {O.}~\bibnamefont {Gühne}}, \bibinfo {author} {\bibfnamefont
  {P.}~\bibnamefont {Hyllus}}, \bibinfo {author} {\bibfnamefont
  {D.}~\bibnamefont {Bruß}}, \bibinfo {author} {\bibfnamefont
  {M.}~\bibnamefont {Lewenstein}}, \ and\ \bibinfo {author} {\bibfnamefont
  {A.}~\bibnamefont {Sanpera}},\ }\href {\doibase
  10.1103/PhysRevLett.92.087902} {\bibfield  {journal} {\bibinfo  {journal}
  {Phys. Rev. Lett.}\ }\textbf {\bibinfo {volume} {92}},\ \bibinfo {pages}
  {087902} (\bibinfo {year} {2004})}\BibitemShut {NoStop}%
\bibitem [{\citenamefont {Woodhead}\ and\ \citenamefont
  {Pironio}(2015)}]{woodhead_secrecy_2015}%
  \BibitemOpen
  \bibfield  {author} {\bibinfo {author} {\bibfnamefont {E.}~\bibnamefont
  {Woodhead}}\ and\ \bibinfo {author} {\bibfnamefont {S.}~\bibnamefont
  {Pironio}},\ }\href {\doibase 10.1103/PhysRevLett.115.150501} {\bibfield
  {journal} {\bibinfo  {journal} {Phys. Rev. Lett.}\ }\textbf {\bibinfo
  {volume} {115}},\ \bibinfo {pages} {150501} (\bibinfo {year}
  {2015})}\BibitemShut {NoStop}%
\bibitem [{\citenamefont {Chaves}\ \emph {et~al.}(2015)\citenamefont {Chaves},
  \citenamefont {Brask},\ and\ \citenamefont
  {Brunner}}]{chaves_device-independent_2015}%
  \BibitemOpen
  \bibfield  {author} {\bibinfo {author} {\bibfnamefont {R.}~\bibnamefont
  {Chaves}}, \bibinfo {author} {\bibfnamefont {J.~B.}\ \bibnamefont {Brask}}, \
  and\ \bibinfo {author} {\bibfnamefont {N.}~\bibnamefont {Brunner}},\ }\href
  {\doibase 10.1103/PhysRevLett.115.110501} {\bibfield  {journal} {\bibinfo
  {journal} {Phys. Rev. Lett.}\ }\textbf {\bibinfo {volume} {115}},\ \bibinfo
  {pages} {110501} (\bibinfo {year} {2015})}\BibitemShut {NoStop}%
\bibitem [{\citenamefont {Himbeeck}\ \emph {et~al.}(2017)\citenamefont
  {Himbeeck}, \citenamefont {Woodhead}, \citenamefont {Cerf}, \citenamefont
  {García-Patrón},\ and\ \citenamefont
  {Pironio}}]{himbeeck_semi-device-independent_2017}%
  \BibitemOpen
  \bibfield  {author} {\bibinfo {author} {\bibfnamefont {T.~V.}\ \bibnamefont
  {Himbeeck}}, \bibinfo {author} {\bibfnamefont {E.}~\bibnamefont {Woodhead}},
  \bibinfo {author} {\bibfnamefont {N.~J.}\ \bibnamefont {Cerf}}, \bibinfo
  {author} {\bibfnamefont {R.}~\bibnamefont {García-Patrón}}, \ and\ \bibinfo
  {author} {\bibfnamefont {S.}~\bibnamefont {Pironio}},\ }\href {\doibase
  10.22331/q-2017-11-18-33} {\bibfield  {journal} {\bibinfo  {journal}
  {Quantum}\ }\textbf {\bibinfo {volume} {1}},\ \bibinfo {pages} {33} (\bibinfo
  {year} {2017})}\BibitemShut {NoStop}%
\bibitem [{\citenamefont {Brask}\ \emph {et~al.}(2017)\citenamefont {Brask},
  \citenamefont {Martin}, \citenamefont {Esposito}, \citenamefont {Houlmann},
  \citenamefont {Bowles}, \citenamefont {Zbinden},\ and\ \citenamefont
  {Brunner}}]{brask_megahertz-rate_2017}%
  \BibitemOpen
  \bibfield  {author} {\bibinfo {author} {\bibfnamefont {J.~B.}\ \bibnamefont
  {Brask}}, \bibinfo {author} {\bibfnamefont {A.}~\bibnamefont {Martin}},
  \bibinfo {author} {\bibfnamefont {W.}~\bibnamefont {Esposito}}, \bibinfo
  {author} {\bibfnamefont {R.}~\bibnamefont {Houlmann}}, \bibinfo {author}
  {\bibfnamefont {J.}~\bibnamefont {Bowles}}, \bibinfo {author} {\bibfnamefont
  {H.}~\bibnamefont {Zbinden}}, \ and\ \bibinfo {author} {\bibfnamefont
  {N.}~\bibnamefont {Brunner}},\ }\href {\doibase
  10.1103/PhysRevApplied.7.054018} {\bibfield  {journal} {\bibinfo  {journal}
  {Phys. Rev. Applied}\ }\textbf {\bibinfo {volume} {7}},\ \bibinfo {pages}
  {054018} (\bibinfo {year} {2017})}\BibitemShut {NoStop}%
\bibitem [{\citenamefont {Wang}\ \emph {et~al.}(2019)\citenamefont {Wang},
  \citenamefont {Primaatmaja}, \citenamefont {Lavie}, \citenamefont
  {Varvitsiotis},\ and\ \citenamefont {Lim}}]{wang_characterising_2019}%
  \BibitemOpen
  \bibfield  {author} {\bibinfo {author} {\bibfnamefont {Y.}~\bibnamefont
  {Wang}}, \bibinfo {author} {\bibfnamefont {I.~W.}\ \bibnamefont
  {Primaatmaja}}, \bibinfo {author} {\bibfnamefont {E.}~\bibnamefont {Lavie}},
  \bibinfo {author} {\bibfnamefont {A.}~\bibnamefont {Varvitsiotis}}, \ and\
  \bibinfo {author} {\bibfnamefont {C.~C.~W.}\ \bibnamefont {Lim}},\ }\href
  {\doibase 10.1038/s41534-019-0133-3} {\bibfield  {journal} {\bibinfo
  {journal} {npj Quantum Information}\ }\textbf {\bibinfo {volume} {5}},\
  \bibinfo {pages} {17} (\bibinfo {year} {2019})}\BibitemShut {NoStop}%
\bibitem [{\citenamefont {Rusca}\ \emph {et~al.}(2019)\citenamefont {Rusca},
  \citenamefont {van Himbeeck}, \citenamefont {Martin}, \citenamefont {Brask},
  \citenamefont {Shi}, \citenamefont {Pironio}, \citenamefont {Brunner},\ and\
  \citenamefont {Zbinden}}]{rusca_practical_2019}%
  \BibitemOpen
  \bibfield  {author} {\bibinfo {author} {\bibfnamefont {D.}~\bibnamefont
  {Rusca}}, \bibinfo {author} {\bibfnamefont {T.}~\bibnamefont {van Himbeeck}},
  \bibinfo {author} {\bibfnamefont {A.}~\bibnamefont {Martin}}, \bibinfo
  {author} {\bibfnamefont {J.~B.}\ \bibnamefont {Brask}}, \bibinfo {author}
  {\bibfnamefont {W.}~\bibnamefont {Shi}}, \bibinfo {author} {\bibfnamefont
  {S.}~\bibnamefont {Pironio}}, \bibinfo {author} {\bibfnamefont
  {N.}~\bibnamefont {Brunner}}, \ and\ \bibinfo {author} {\bibfnamefont
  {H.}~\bibnamefont {Zbinden}},\ }\href {http://arxiv.org/abs/1904.04819}
  {\bibfield  {journal} {\bibinfo  {journal} {arXiv:1904.04819 [quant-ph]}\ }
  (\bibinfo {year} {2019})}\BibitemShut {NoStop}%
\bibitem [{\citenamefont {Ivanovic}(1987)}]{ivanovic_how_1987}%
  \BibitemOpen
  \bibfield  {author} {\bibinfo {author} {\bibfnamefont {I.~D.}\ \bibnamefont
  {Ivanovic}},\ }\href@noop {} {\bibfield  {journal} {\bibinfo  {journal}
  {Phys. Lett. A}\ }\textbf {\bibinfo {volume} {123}},\ \bibinfo {pages} {3}
  (\bibinfo {year} {1987})}\BibitemShut {NoStop}%
\bibitem [{\citenamefont {Dieks}(1988)}]{dieks_overlap_1988}%
  \BibitemOpen
  \bibfield  {author} {\bibinfo {author} {\bibfnamefont {D.}~\bibnamefont
  {Dieks}},\ }\href {\doibase 10.1016/0375-9601(88)90840-7} {\bibfield
  {journal} {\bibinfo  {journal} {Phys. Lett. A}\ }\textbf {\bibinfo {volume}
  {126}},\ \bibinfo {pages} {303} (\bibinfo {year} {1988})}\BibitemShut
  {NoStop}%
\bibitem [{\citenamefont {Peres}(1988)}]{peres_how_1988}%
  \BibitemOpen
  \bibfield  {author} {\bibinfo {author} {\bibfnamefont {A.}~\bibnamefont
  {Peres}},\ }\href {\doibase 10.1016/0375-9601(88)91034-1} {\bibfield
  {journal} {\bibinfo  {journal} {Phys. Lett. A}\ }\textbf {\bibinfo {volume}
  {128}},\ \bibinfo {pages} {19} (\bibinfo {year} {1988})}\BibitemShut
  {NoStop}%
\bibitem [{\citenamefont {Uhlmann}(1976)}]{uhlmann_transition_1976}%
  \BibitemOpen
  \bibfield  {author} {\bibinfo {author} {\bibfnamefont {A.}~\bibnamefont
  {Uhlmann}},\ }\href {\doibase 10.1016/0034-4877(76)90060-4} {\bibfield
  {journal} {\bibinfo  {journal} {Reports on Mathematical Physics}\ }\textbf
  {\bibinfo {volume} {9}},\ \bibinfo {pages} {273} (\bibinfo {year}
  {1976})}\BibitemShut {NoStop}%
\bibitem [{\citenamefont {Jozsa}(1994)}]{jozsa_fidelity_1994}%
  \BibitemOpen
  \bibfield  {author} {\bibinfo {author} {\bibfnamefont {R.}~\bibnamefont
  {Jozsa}},\ }\href {\doibase 10.1080/09500349414552171} {\bibfield  {journal}
  {\bibinfo  {journal} {Journal of Modern Optics}\ }\textbf {\bibinfo {volume}
  {41}},\ \bibinfo {pages} {2315} (\bibinfo {year} {1994})}\BibitemShut
  {NoStop}%
\bibitem [{\citenamefont {Nielsen}\ and\ \citenamefont
  {Chuang}(2011)}]{Nielsen_QCQ_2011}%
  \BibitemOpen
  \bibfield  {author} {\bibinfo {author} {\bibfnamefont {M.~A.}\ \bibnamefont
  {Nielsen}}\ and\ \bibinfo {author} {\bibfnamefont {I.~L.}\ \bibnamefont
  {Chuang}},\ }\href@noop {} {\emph {\bibinfo {title} {Quantum Computation and
  Quantum Information: 10th Anniversary Edition}}},\ \bibinfo {edition} {10th}\
  ed.\ (\bibinfo  {publisher} {Cambridge University Press},\ \bibinfo {address}
  {New York, NY, USA},\ \bibinfo {year} {2011})\BibitemShut {NoStop}%
\bibitem [{\citenamefont {Ioannou}\ \emph {et~al.}(2019)\citenamefont
  {Ioannou}, \citenamefont {Brask},\ and\ \citenamefont
  {Brunner}}]{ioannou_upper_2019}%
  \BibitemOpen
  \bibfield  {author} {\bibinfo {author} {\bibfnamefont {M.}~\bibnamefont
  {Ioannou}}, \bibinfo {author} {\bibfnamefont {J.~B.}\ \bibnamefont {Brask}},
  \ and\ \bibinfo {author} {\bibfnamefont {N.}~\bibnamefont {Brunner}},\ }\href
  {\doibase 10.1103/PhysRevA.99.052338} {\bibfield  {journal} {\bibinfo
  {journal} {Phys. Rev. A}\ }\textbf {\bibinfo {volume} {99}},\ \bibinfo
  {pages} {052338} (\bibinfo {year} {2019})}\BibitemShut {NoStop}%
\end{thebibliography}%



\end{document}